\documentclass[aps,aps,twocolumn,showpacs,preprintnumbers,amsmath,amssymb]{revtex4}
\usepackage{lipsum}
\usepackage{graphicx}% Include figure files
\usepackage[latin1]{inputenc}
\usepackage{dcolumn}% Align table columns on decimal point
\usepackage{bm}% bold math
\usepackage{epsfig}
\usepackage{color}

\begin{document}

%\preprint{APS/123-QED}

\title{Sequential visibility graph motifs}% Force line breaks with \\

\author{Jacopo Iacovacci, Lucas Lacasa}
\email{j.iacovacci@qmul.ac.uk, l.lacasa@qmul.ac.uk}
\affiliation{School of Mathematical Sciences, Queen Mary University of London, Mile End Road, E14NS London (UK)}%

\date{\today}% It is always \today, today,
             %  but any date may be explicitly specified

\begin{abstract}
Visibility algorithms transform time series into graphs and encode dynamical information in their topology, paving the way for graph-theoretical time series analysis as well as building a bridge between nonlinear dynamics and network science. In this work we introduce and study the concept of sequential visibility graph motifs, smaller substructures of $n$ consecutive nodes that appear with characteristic frequencies. We develop a theory to compute in an exact way the motif profiles associated to general classes of deterministic and stochastic dynamics. We find that this simple property is indeed a highly informative and computationally efficient feature capable to distinguish among different dynamics and robust against noise contamination. We finally confirm that it can be used in practice to perform unsupervised learning, by extracting motif profiles from experimental heart-rate series and being able, accordingly, to disentangle meditative from other relaxation states. Applications of this general theory include the automatic classification and description of physical, biological, and financial time series.
\end{abstract}

\pacs{}% PACS, the Physics and Astronomy
                             % Classification Scheme.
\keywords{} \maketitle

\section{Introduction}

The interdisciplinary field of Network Science \cite{barabasirev,newman2003,Boccaletti2006, Newman2010} has integrated in the last 15 years under a single paradigm tools and techniques coming from the Mathematics (Combinatorics and Graph Theory), Physics (Statistical Physics) and Computer Science (Machine Learning and Data Mining) communities, in the task of exploring, characterizing and modelling the structure and function of large and complex networks arising in nature, technology and society. Perhaps one of the most interesting concepts that has emerged within this synergy is that of network motifs, small subgraphs appearing with statistically significant frequencies that are suggested to represent building blocks of network architecture \cite{uri2}. This local topological feature has proved to be very useful for classifying large graphs in areas as biochemistry, neuroscience or ecology to cite a few (for instance networks that process information of any garment seem to share similar motif statistics \cite{uri2}), or for understanding the interplay between network's local structure and function \cite{uri}. One can even use the local information gathered by motif statistics to compare networks of different sizes, enabling a classification of networks in terms of superfamilies \cite{uri3}. Both the role played by network motifs as well as the computational problem of efficiently extracting network motifs \cite{motif_extraction} are two areas of current active research.\\
Of course, this useful structural descriptor -and in general any topological measure- is narrowed down to those datasets and systems that have a natural representation in terms of graphs. As a matter of fact, in some of the most challenging and complex systems that scientists face nowadays (let it be spatio-temporal chaotic, or turbulent systems, the financial system, brain activity, etc), information is available in the form of temporal streams of data: series describing the time evolution of certain observables.  Interestingly enough, in recent years a novel branch in data analysis has started to transform time series into graph-theoretical representations. Among other interesting possibilities \cite{Small, Thurner2007, Small2, donner2010, donner2011}, the family of visibility algorithms \cite{lacasa2008from, pre, epjb, multivariate} stand out as computationally simple methods to transform time series into networks which are capable of mapping seemingly hidden structure of the series and the underlying dynamics into graph space, with the peculiarity of often being analytically tractable \cite{lacasa2008from}. Here we extend, via visibility algorithms,  a tailored notion of network motifs to the realm of time series analysis and classification \cite{Small2}.\\

\noindent The rest of the paper goes as follows. After recalling the basics of visibility (VG) and horizontal visibility graphs (HVG), we define sequential VG/HVG motifs (Section II) and develop a mathematical theory for the HVG case  (Section III) that allows us to easily derive analytical expressions for the motif profiles of several classes of stochastic and deterministic dynamical systems. We prove, accordingly, that sequential HVG motifs are informative features that can easily distinguish among different types of complex dynamics. In section IV we further show that such discrimination is robust, even when the signals under study are polluted by large amounts of measurement noise, enabling its use in empirical (experimental) time series, i.e. in practical problems. We summarise our results on synthetic time series in section V and finally make use of this methodology in a real scenario in section VI, where we are able classify different physiological time series and efficiently disentangle meditative from general relaxation states by using the motif profiles (only five numbers per subject) extracted from heartbeat time series. In section VII we conclude.

\section{Visibility graphs and motifs}
\begin{figure*}
\includegraphics[width= 16 cm]{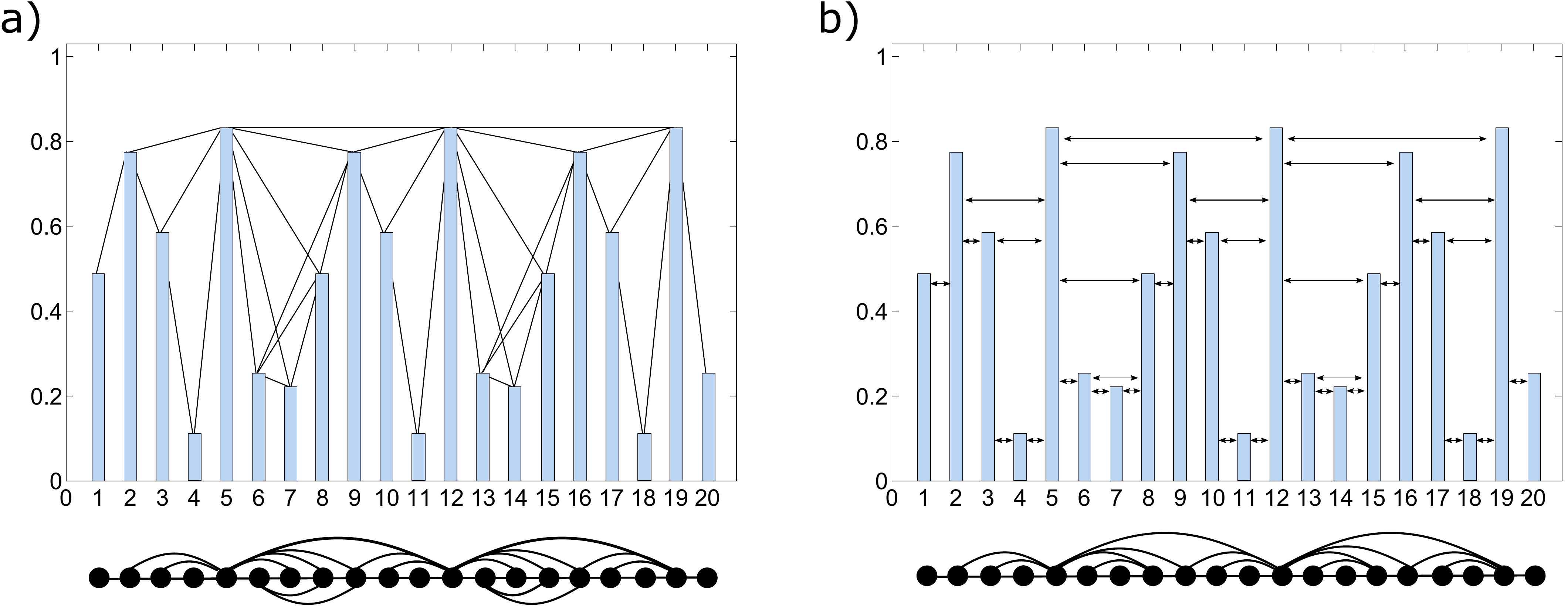}
\caption{(Color online) Schematic of two families of visibility algorithms. (a): Natural Visibility Algorithm 
applied to 20 data points of a periodic time series (top) and the corresponding Visibility Graph (VG) (bottom); each
datum in the series corresponds to a node in the graph and two nodes are connected if their corresponding data heights show mutual visibility (see the text). (b): Horizontal Visibility Algorithm applied to the same series (top) and the corresponding Horizontal Visibility Graph (HVG) (bottom); each datum in the series corresponds to a node in the graph and two nodes are connected if their corresponding data heights show horizontal visibility (see the text).}
\label{fig:0}
\end{figure*}

Visibility algorithms \cite{lacasa2008from, pre, epjb, multivariate} are a family of methods to map time series into graphs, in order to exploit the tools of graph theory and network science to describe and characterise both the structure of time series and their underlying dynamics. Let ${\cal S}=\{x(t)\}_{t=1}^T$ be a real-valued time series of $T$ data. A so called natural visibility graph (VG) is a planar graph of $T$ nodes in association to $\cal S$, such that (i) every datum $x(i)$ in the series is related to a node $i$ in the graph (hence the graph nodes inherit a natural ordering), and (ii) two nodes $i$ and $j$ are connected by an edge if any other datum $x(k)$ where $i<k<j$ fulfils the following \textit{convexity} criterion:
$$x_k< x_i + \frac{k-i}{j-i}[x_j-x_i],\ \forall k: i<k<j$$
By construction, VGs are connected graphs with a natural Hamiltonian path given by the sequence of nodes $(1,2,\dots,T)$, whose topology is invariant under a set of basic transformations in the series, including horizontal and vertical translations. An illustration of this method is shown in panel (a) of figure \ref{fig:0}, where we plot a time series and its associated VG. VGs inherit in its topology the structure of the time series, in such a way that periodic, random, and fractal
series map into motif-like, random exponential and scale-free networks, respectively. It has been shown that VGs are well suited to handle non-stationary data \cite{epl, RW_visibility, RW_visibility2}.\\

\noindent A so called horizontal visibility graph (HVG) is defined as a subgraph of the VG, obtained by restricting the visibility criterion and imposing horizontal visibility instead. In this case, two nodes $i$ and $j$ are connected by an edge in the HVG if any other datum $x(k)$ where $i<k<j$ fulfil the following \textit{ordering} criterion:
 $$x_k<\inf(x_i,x_j),\ \forall k: i<k<j$$
Such subgraph is indeed an outerplanar graph \cite{severini} (see Figure \ref{fig:0}, panel b) for an illustration). Interestingly, HVG inherits some of the properties of VGs and, on top of that, are computationally more efficient \footnote{According to numerical experiments, HVG has linear complexity on aperiodic dynamics $\text{http://www.maths.qmul.ac.uk/~lacasa/Software.html}$} and analytically tractable. Accordingly, several analytical properties of these family of graphs \cite{lacasa2014on, pre}, associated to different classes of dynamics including canonical routes to chaos \cite{plosone, pre2013, jpa2014, jns} have been investigated in recent years. For instance, for the class of Markovian processes with an integrable invariant measure the values of the degree distribution
$P(k)$ can be calculated analytically using a formal diagrammatic theory \cite{lacasa2014on}.\\
 
 \noindent We are now ready to introduce a new topological property of VG/HVG.\\

\noindent {\bf Definition} \textit{(sequential VG/HVG $n$-node motifs)}. Consider a VG/HVG of $N$ nodes, associated to a time series of $N$ data, and label the nodes according to the natural ordering induced by the arrow of time (i.e. the trivial Hamiltonian path). Set $n<N$ and consider, sequentially, all the subgraphs formed by the sequence of nodes $\{s,s+1,\dots,s+n-1\}$ (where $s$ is an integer that takes values in $[1,N-n+1]$) and the edges from the VG/HVG only connecting these nodes: these are defined as the sequential $n$-node motifs of the VG/HVG. This is akin to defining a sliding window of size $n$ in graph space that initially covers the first $n$ nodes and sequentially slides, in such a way that for each window, one can associate a motif by (only) considering the edges between the $n$ nodes belonging to that window.\\

\noindent Note that, importantly, this definition differs from the one of a standard network motif (which looks at the frequencies of appearance of all subgraphs of a given size, without imposing any restriction on the nodes forming a given subgraph), as here it is required that the labels of the nodes appearing in a motif are in strict sequential order -this is consistent with the vertex ordering of the natural Hamiltonian path induced by construction in the VGs/HVGs-. That is, in order to preserve in graph space the dynamical information of the series, the $n$ nodes of an $n$-size motif are taken in sequential order, and only those edges that connect nodes from the motif are considered. For readability, from now on we will call these simply VG/HVG motifs but the reader should not get confused and remind that these are not directly the standard notion of network motifs computed on a VG/HVG. Some basic properties of these motifs are:
\begin{itemize}
\item Trivially, there is a total of $N-n$ motifs (which can be the same motifs or not) within each VG/HVG. 
\item Each motif is a subgraph of the original VG/HVG. Moreover, HVG motifs are outerplanar and have a trivial Hamiltonion path, thus HVG motifs are also HVGs \cite{severini}. As a result, there are only 6 admissible motifs of size 4, and 2 admissible motifs of size 3 (see table \ref{ineq4} for an enumeration).
\item Computational complexity: Computing motifs in both VG and HVG is extremely efficient. If instead of exploring the motif occurrence in the structure of the adjacency matrix, one directly examines the set of inequalities reported in table  \ref{ineq4}, one directly has an algorithm that runs in \textit{linear} time $O(N)$ for HVG motifs. A similar complexity is found for VG motifs \cite{inprep}.
\end{itemize}  

\noindent As is done traditionally with network motifs \cite{uri3}, we can compare VG/HVGs associated to different time series and dynamics by comparing the relative occurrence of each motif inside a VG/HVG. In order to do that, we introduce the extension to the VG/HVG realm of a significance profile:\\ 

\noindent {\bf Definition} \textit{(VG/HVG motif profile ${\bf Z}^n$)}. Let $p$ be the total number of admissible VG/HVG motifs with $n$-nodes. Assign to each of these $p$ motifs a label from $1$ to $p$ (that is, choose an ordering for the motifs). The motif assigned with the label $i$ will be called a type-$i$ motif.
Then, we define the $n$-node VG/HVG motif significance profile ${\bf Z}^n$ (or simply HVG motif profile) of a certain time series of size $N$ as the vector function ${\bf Z}^n: n\in \mathbb{N}\to [\mathbb{P}^n_1,\dots,\mathbb{P}^n_p] \in [0,1]^p$ whose output is a vector of $p$ components, where the $i$-th component, $\mathbb{P}^n_i$, is the relative frequency of the type-$i$ motif.\\

\noindent Several technical comments are in order: 
\begin{itemize}
\item First, since ${\bf Z}^n$ are $n$-dimensional real vectors, any $L_p$ norm induces a natural similarity measure (distance) between two graphs.
\item Second, ${\bf Z}^n$ has, by construction, unit $L_1$-norm, as $\sum_{i=1}^p |\mathbb{P}^n_i|=\sum_{i=1}^p\mathbb{P}^n_i=1$.
\item Third, note that if one considers dynamical processes instead of individual time series, then the estimated relative frequencies $\mathbb{P}^n_i$ for an individual realization of the dynamical process converge for infinitely long series to the probabilities of type-$i$ motif associated to the process. For the motif profile to be a well-defined feature of a certain dynamical process, it needs to be self-averaging. We check this property by estimating ${\bf Z}^n$ for an ensemble of realizations of the process, computing the mean $\langle \mathbb{P}^n_i \rangle$ and standard deviation $\sqrt{\langle [\mathbb{P}^n_i]^2 \rangle-\langle \mathbb{P}^n_i \rangle^2}$ over this ensemble, and checking that the standard deviation is small (meaning that a single realization provides a good description of the average behaviour). As we will show below, both VG and HVG motif profiles have very good self-averaging properties. In any case, for every dynamical process considered in this work, instead of $\mathbb{P}^n_i$ we compute  $\langle \mathbb{P}^n_i \rangle$ and $\sqrt{\langle [\mathbb{P}^n_i]^2 \rangle-\langle \mathbb{P}^n_i \rangle^2}$, but for readability, from now on we will drop the $\langle \cdot \rangle$ for the elements of the motif profile, as we found that for the size of the series used in the numerical analysis, $\sqrt{\langle [\mathbb{P}^n_i]^2 \rangle-\langle \mathbb{P}^n_i \rangle^2}$ was very small and hence $\mathbb{P}^n_i \approx \langle \mathbb{P}^n_i \rangle$.
\item Fourth, note at this point that the definition of the VG/HVG motif profile is different from standard profiles (significance profile, subgraph ratio profile) defined in the literature \cite{uri3}, as in the latter case, they make use of a null model (ensemble of randomised networks) to appropriately normalise each frequency. The rationale for this normalization is that one wants to compare motif statistics across very different networks (with different sizes and degree sequences), so variations in the motif relative frequencies only due to size effects need to be removed to be able to correctly compare across different networks. The reader will quickly come to the conclusion that, in the context of VG/HVG, the null model is not a randomised ensemble of the graph under study (which would not yield a VG/HVG with high probability), but on the contrary, it should be the VG/HVG of a randomisation of the \textit{time series} under study. In other words, normalisation in the case of VG/HVG profiles should deal with the motif statistics of uncorrelated random series (i.i.d. white noise or surrogate series that preserve certain structures) with similar probability densities than the series under study. In the next section we will prove that, in the case of HVGs (which will be the family of visibility graphs under study), such null model has a universal motif profile, independent of the probability density of the i.i.d. process. Therefore, it is not necessary in this case to normalise each profile accordingly as this would only yield a trivial, constant rescaling.
\end{itemize}
\begin{figure}
\includegraphics[width= 7 cm]{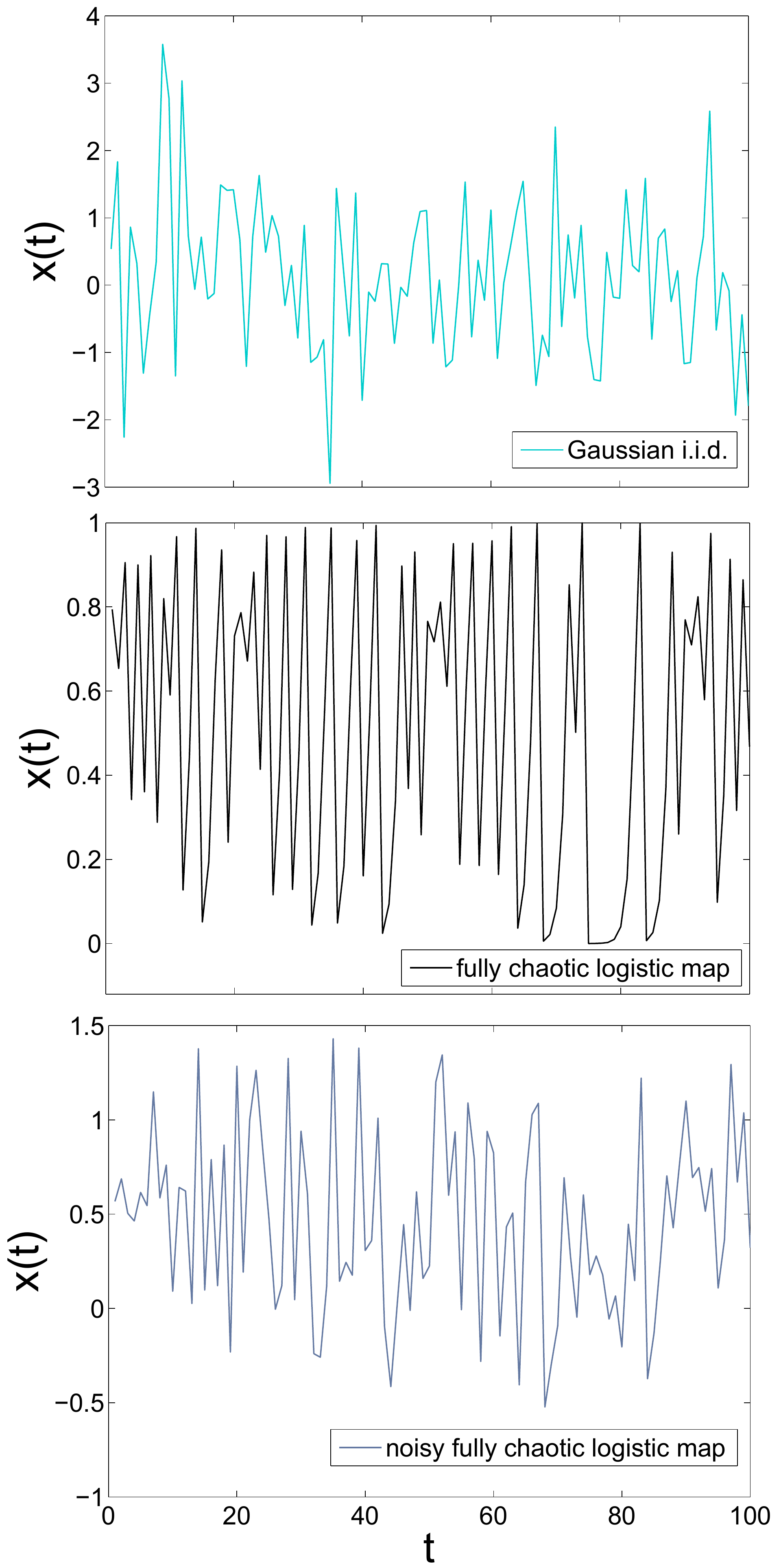}
\caption{(Color online)Sample time series from (a) i.i.d. Gaussian white noise, (b) fully chaotic logistic map, and (c) fully chaotic logistic map polluted with a certain amount of extrinsic white noise are shown for illustrative purpose. Visibility graph motifs can be extracted from these series to reveal differences in their intrinsic structure.}
\label{series}
\end{figure}

\begin{figure}
\centering
\includegraphics[width= 8.5 cm]{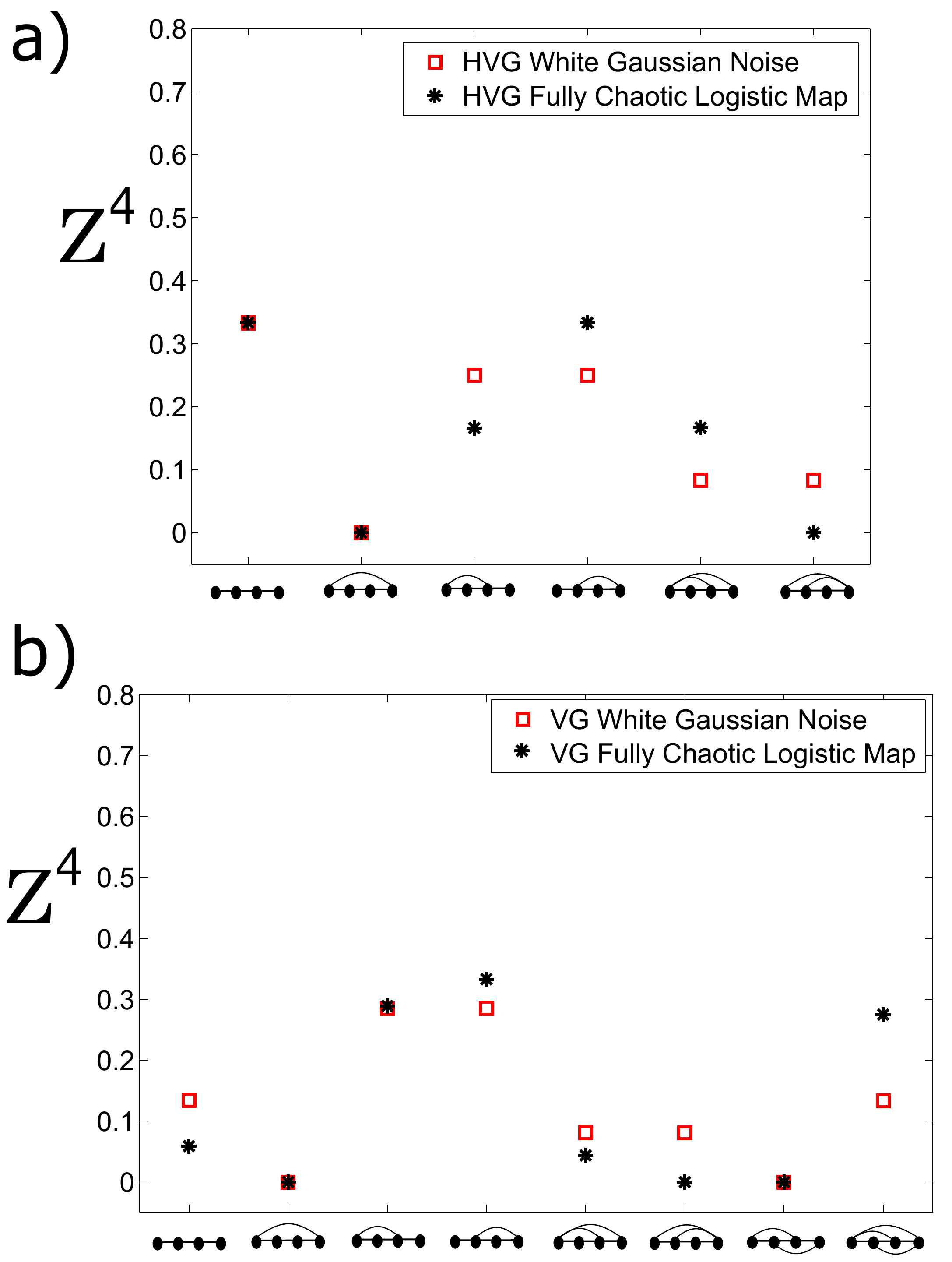}
\caption{(Color online) $4$-node motif profiles ${\bf Z}^4$ associated to Gaussian white noise (red squares) and to a fully chaotic logistic map (black stars) extracted respectively from HVG (panel a) and VG (panel b). Each dot represents the relative frequency of a given motif, averaged over an ensemble of 100 realizations of each process (time series of $N=10^4$ data per realization). Standard deviations of each motif relative frequency over the ensemble are plotted as error bars, which are not visible as error bars fall inside the symbols. 
We conclude that these motifs can be used to distinguish between deterministic and stochastic dynamics.}
\label{fig:1}
\end{figure}

\noindent For illustration purposes, let $n=4$, and consider two different dynamical processes: (i) white Gaussian noise described by the map $x_t= \xi$, where $\xi$ are independent and identically distributed (i.i.d.) Gaussian random variables $\xi \sim {\cal N}[0,1]$, and (ii) chaotic dynamics given by the fully chaotic logistic map $x_{t+1}= 4x_{t}(1-x_{t})$. 
In order to estimate the probability of appearance of each of the motifs, we have generated a time series of size $N=10^4$ data for both processes (sample time series can be seen in the top panels of figure \ref{series}), and we have computed the relative frequencies of each motif. Results, averaged over an ensemble of 100 realizations, are shown in figure \ref{fig:1} (error bars describing the ensemble standard deviation are contained inside the symbols); in panel (a) we plot the HVG motif profile, whereas in panel (b) we plot the VG profile. As we can see, in every case the type-II motif is absent. The simple reason is that this profile is absent for irregular (aperiodic) real-valued time series, by construction (see table \ref{ineq4}). 

For the chaotic process, some other motifs are absent: this is related to forbidden patterns arising in chaotic dynamics. More importantly, in both panels, the average relative frequency of some motifs seems to be different for both dynamical processes, enabling the possibility of using both HVG and VG motif profiles to distinguish amongst different dynamical origins. 
From now on we will focus our motif analysis on the horizontal visibility graphs (HVG) alone, and comparison with the VG case is left for future work \cite{inprep}. In the next section we advance a theory to compute the motif profile ${\bf Z}^n$ in an exact way for different classes of dynamical systems. We will confirm that HVG motifs can indeed distinguish several kinds of dynamics, and we will explore how to build on this peculiar property for feature-based classification.

\section{Theory}
\label{sec:theory}
In order to numerically explore and compute the frequency of each HVG motif, one can generate the HVG associated to a given time series and count the presence of each motif directly from the adjacency matrix. However, in this section we will show that it is not necessary to do that as, via the zero-order terms of a diagrammatic expansion recently advanced \cite{lacasa2014on}, we can also work out the motif occurrence directly from the exploration of the time series, that enables motif computation in linear time. This will allow us to build a theory by which the motif profiles can be computed exactly for a large set of classes of dynamics that fulfil certain properties. Let us consider a dynamical process ${\cal H}:\mathbb{R}\to \mathbb{R}$ with a smooth invariant measure $f(x)$ that fulfils the Markov property. That is, from a probabilistic point of view, conditional probabilities fulfil $f(x_n|x_{n-1},x_{n-2},\dots)=f(x_n|x_{n-1})$, where $f(x_n|x_{n-1})$ is the transition probability distribution (note that this concept has a clear meaning in random dynamical systems, whereas for deterministic systems, say maps $x_{t+1}={\cal H}(x_t)$, the Markov property is also trivially fulfilled with $f(x_2|x_1)=\delta(x_2-{\cal H}(x_1))$, where $\delta(x)$ is the Dirac-delta distribution).
 The key element is that for these processes, each HVG motif has a probability of appearance as a subgraph that can \textit{directly} be computed as the measure of a \textit{set of ordering inequalities} that take place in the time series. For instance, for $n=3$ and $n=4$, probabilities associated to the appearance of a certain motif are based on integrals of the form:
 \begin{equation}
\int f(x_0) dx_0 \int f(x_1|x_0) dx_1 \int f(x_2|x_1) dx_2
\label{gen3}
\end{equation}
for $n=3$, and 
\begin{equation}
\int f(x_0) dx_0 \int f(x_1|x_0) dx_1 \int f(x_2|x_1) dx_2 \int f(x_3|x_2) dx_3
\label{gen4}
\end{equation}
for $n=4$.
\begin{table*}
%\begin{ruledtabular}
\begin{tabular}{ccc}
{\bf Motif label}&{\bf Motif type}&{\bf Inequality set}\\
\hline
1&\begin{minipage}{.1\textwidth}
\includegraphics[width= 0.5\textwidth]{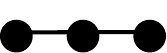}
\end{minipage}
&$\{ \forall (x_0,x_2), x_1>x_0 \}\cup\{ \forall x_0, x_1<x_0,x_2<x_1\} $\\
2&\begin{minipage}{.1\textwidth}
\includegraphics[width= 0.5\textwidth]{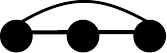}
\end{minipage}& $\{\forall x_0, x_1<x_0, x_2>x_1\}$\\
\hline
1&\begin{minipage}{.1\textwidth}
\includegraphics[width= 0.7\textwidth]{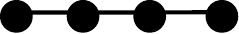}
\end{minipage}&$\{ \forall (x_0,x_1), x_2<x_1, x_3<x_2\}\cup\{\forall (x_0,x_3), x_1>x_0,x_2>x_1\} $\\
2&\begin{minipage}{.1\textwidth}
\includegraphics[width= 0.7\textwidth]{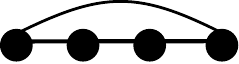}
\end{minipage}& $\{\forall x_0, x_1<x_0,x_2=x_1,x_3>x_2\}$ \\
3&\begin{minipage}{.1\textwidth}
\includegraphics[width= 0.7\textwidth]{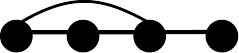}
\end{minipage}& $\{\forall x_0, x_1<x_0, x_1<x_2<x_0,x_3<x_2\}\cup\{\forall (x_0,x_3),x_1<x_0,x_2>x_0\}$\\
4&\begin{minipage}{.1\textwidth}
\includegraphics[width= 0.7\textwidth]{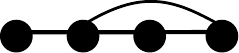}
\end{minipage}& $\{ \forall x_0, x_1>x_0,x_2<x_1, x_3>x_2  \}\cup\{\forall x_0, x_1<x_0, x_2<x_1, x_2<x_3<x_1\}$\\
5&\begin{minipage}{.1\textwidth}
\includegraphics[width= 0.7\textwidth]{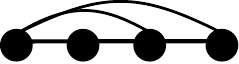}
\end{minipage}&$\{\forall x_0, x_1<x_0, x_1<x_2<x_0, x_3>x_2\}$\\
6&\begin{minipage}{.1\textwidth}
\includegraphics[width= 0.7\textwidth]{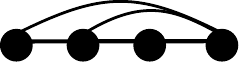}
\end{minipage}& $\{\forall x_0, x_1<x_0, x_2<x_1, x_3>x_1\}$\\
\end{tabular}
%\end{ruledtabular}
\caption{Enumeration of all $3$ and $4$-node motifs. Each motif can be characterized according to a hierarchy of inequalities in the associated time series. Note that for real-valued aperiodic dynamics the type-II $4$-node motif has a null probability of occurrence as the probability that two data in the time series repeat vanishes almost surely (if, on the other hand, the series only take values from a finite set then this motif has a finite probability). For the rest, the probability of each motif reduces to the measure of the set of inequalities (see the text).}
	\label{ineq4}
\end{table*}
The range of integration and the shape of the conditional probabilities are particular for each motif and each process, respectively. First, the range of integration fully determines the motif. In table \ref{ineq4} we depict the conditions in the time series that have to be fulfilled among $n$ consecutive data $x_0,x_1,\dots,x_{n-1}$ to yield a certain motif of size $n$ in the HVG, for $n=3,4$ (extension to arbitrary $n$ is easy but gets cumbersome as $n$ increases). It can be proved quite easily that a given motif appears in an HVG if and only if these ordering restrictions are fulfilled in the time series. These restrictions directly translate in the integration range of the probabilities, we illustrate this principle in an example. The first motif, ${\bf Z}_1^4$,  according to table \ref{ineq4} is guaranteed when 4 consecutive values $x_0,x_1,x_2$ and $x_3$ are such that $\{ \forall (x_0,x_1), x_2<x_1, x_3<x_2\}\cup\{\forall (x_0,x_3), x_1>x_0,x_2>x_1\} $. Accordingly, if $x \in [a,b] \subset \mathbb{R}$, the probability of this event is
\begin{widetext}
\begin{align}
{\bf Z}_1^4\equiv\mathbb{P}^4_1=\int_a^b f(x_0) dx_0& \int_a^b f(x_1|x_0) dx_1 \int_a^{x_1} f(x_2|x_1) dx_2 \int_a^{x_2} f(x_3|x_2) dx_3 + \nonumber\\
&\int_a^b f(x_0) dx_0 \int_{x_0}^b f(x_1|x_0) dx_1 \int_{x_1}^b f(x_2|x_1) dx_2 \int_a^b f(x_3|x_2) dx_3 
\label{ex1}.
\end{align}
\end{widetext}
Analogous expressions can be found for the rest of the probabilities that form the motif profile $\bf{Z}$. These terms are nothing but the contributions to the degree distribution at zero-order from a diagrammatic expansion in the number of hidden nodes \cite{lacasa2014on}. From a geometric point of view, the first motif will not appear in fast fluctuating signals and hence deals with the degree of smoothness of a time series at short (order $n$) scales, whereas the other motifs deal with certain fluctuation shapes. Accordingly, in those processes where the degree of smoothness can vary -such as in fractional Brownian motion, where the smoothness of the signal increases with the Hurst exponent- we would expect that the first motif is particularly informative, whereas for fast-fluctuating series we expect this motif to be less informative. Integrals accounting for the probabilities are easy to deal with; in several cases these are exactly solvable, and in general one can solve them up to arbitrary precision with any symbolic programming software. In what follows we determine the motif profiles for i.i.d. (white noise), coloured noise with exponentially decaying correlations, and deterministic chaos (fully chaotic logistic map). We show that ${\bf Z}^4$ capture enough information to easily distinguish different processes and thus represent excellent features for series classification.\\

\noindent  \textbf{Relation with ordinal patterns.}
At this point it is important to highlight the relation between the probability of occurrence of a given HVG motifs and the probability of occurrence of so called ordinal patterns \cite{PE1, PEbook}. In the theory proposed by Bandt and Pompe in\cite{PE1} for the case of the embedding dimension equal to 4 one proceeds to map each local time series segment of size 4 into an ordering symbol of 4 letters from the alphabet $\{0,1,2,3\}$ (where the largest value maps to the letter $0$, the second largest to $1$, the third largest to $2$, and the smallest to $3$). There are $4!=24$ permutations, defining 24 symbols (ordinal patterns) whose frequencies are then counted to measure the so-called permutation entropy that acts as a complexity measure of the series \cite{PE1}. Interestingly, the probability of occurrence of each HVG motif indeed reduces to the probability of occurrence of a set of possible ordinal patterns (this is no longer the case for VG motifs \cite{inprep}). For instance, ${\bf Z}_1^4$ is the probability of finding any of the ordinal patterns $0123, 1023, 1203, 1230, 2103, 2130, 2310,$ or $3210$, and similarly the rest of the motif probabilities can be linked to the probability of appearance of different sets of ordinal patterns. Accordingly, HVG motifs indeed induce a particular partition of the set of ordinal patterns. The HVG motif profile is thus intimately linked with the so called permutation spectrum \cite{PE3} that accounts for the histogram of ordinal patterns.

\subsection{i.i.d.}
Let us start by considering time series generate by i.i.d. uniform random variables $\xi \sim U[0,1]$. In this case we have $a=0, b=1, f(x)=1$ and $f(x|y)=f(x)\  \forall y$, and simply enough, probabilities defined by eqs. \ref{gen3} and \ref{gen4} easily factorize. According to table \ref{ineq4}, after a little bit of calculus we find 
\begin{eqnarray}
{\bf Z}^3=\bigg[\frac{2}{3},\frac{1}{3}\bigg]; \ {\bf Z}^4=\bigg[\frac{8}{24},0,\frac{6}{24},\frac{6}{24}, \frac{2}{24}, \frac{2}{24}\bigg] 
\label{unif}
\end{eqnarray} 
Note that these results are in perfect quantitative agreement with numerics performed for finite size series (left panel of figure \ref{fig:1}); we will show in the next subsection that results for finite series converge quite fast to the (asymptotic) theory as the series size increases. Interestingly, results indeed coincide despite the fact that the theoretical values were computed for \textit{uniform} white noise ($f(x)=1$), while the numerics in figure \ref{fig:1} were performed on \textit{Gaussian} white noise (where $f(\cdot)$ is the Gaussian function). This suggests that i.i.d. may have a universal HVG motif profile, indeed independent of $f(\cdot)$. We now state and prove a theorem that actually guarantees this result.\\

\noindent {\bf Theorem 1.} Consider a bi-infinite series of i.i.d. random variables extracted from a continuous distribution $f(x)$ with support $(a,b)$, where $a,b \in \mathbb{R}$. Then the probability of finding $n$-node HVG motifs (with $n=3,4$) follows eq. \ref{unif}, independently of the shape of $f(x)$.\\

\noindent {\bf Proof. } The proof is a constructive one. We only give here the explicit proof for $\mathbb{P}^4_1$, as the proof for the rest of probabilities follow analogously. We rely on the cumulative distribution function $F(x)$, defined as $\int_a^x f(x')dx'=F(x)$, with properties $F(a)=0, F(b)=1$ and 
\begin{equation}
f(x)F^{n-1}(x)=\frac{dF^n(x)}{ndx}.
\label{prop}
\end{equation}
We have
\begin{align} 
&\mathbb{P}^4_1= \nonumber \\
& \int_a^b f(x_0) dx_0 \int_a^b f(x_1) dx_1 \int_a^{x_1} f(x_2) dx_2 \int_a^{x_2} f(x_3) dx_3 
+ \nonumber \\ &\int_a^b f(x_0) dx_0 \int_{x_0}^b f(x_1) dx_1 \int_{x_1}^b f(x_2) dx_2 \int_a^b f(x_3) dx_3 \nonumber
\end{align}
Using the properties of $F(x)$, the first term above is then
\begin{align}
&\int_a^b f(x_0) dx_0 \int_a^b f(x_1) dx_1 \int_a^{x_1} f(x_2) dx_2 \int_a^{x_2} f(x_3) dx_3  =\nonumber \\ 
& \int_a^b f(x_0) dx_0 \int_a^b f(x_1) dx_1 \int_a^{x_1} f(x_2)F(x_2) dx_2 =\nonumber\\ 
&\int_a^b f(x_0) dx_0 \int_a^b f(x_1) \frac{F^2(x_1)}{2} dx_1= \nonumber\\ 
&\int_a^b \frac{f(x_0)}{6} dx_0=\frac{1}{6}\nonumber, 
\end{align}
and analogously for the second term,
\begin{align}
&\int_a^b f(x_0) dx_0 \int_{x_0}^b f(x_1) dx_1 \int_{x_1}^b f(x_2) dx_2 \int_a^b f(x_3) dx_3 = \nonumber \\
&\int_a^b f(x_0) dx_0 \int_{x_0}^b f(x_1)(1-F(x_1)) dx_1 =\nonumber \\ 
& \int_a^b f(x_0)\bigg[\frac{1}{2}-F(x_0)+\frac{F^2(x_0)}{2}\bigg]dx_0 = \nonumber \\
&\frac{F(x_0)}{2}-\frac{F^2(x_0)}{2}+\frac{F^3(x_0)}{6}\bigg |_a^b=\frac{1}{6},
\end{align}
hence $\mathbb{P}^4_1=2/6=8/24$, coinciding with the result for uniform and Gaussian series, and being independent of $f(x)$. The rest of the elements in ${\bf Z}^4$ are computed analogously.
 $\blacksquare$\\

\noindent As a matter of fact, the independency from $f(x)$ can be trivially extended for an arbitrary size of the motif $n$. This is intuitive so we only give here the strategy of a proof. The main ingredient which is required for this independency to hold $\forall n$ is that the limits of the $n$-th integral are either the extremes of the distribution support $a,b$ (where the cumulative distribution $F(x)$ take the constant values 0 and 1 respectively, and independently of $f(x)$), or other variables $x_0 \dots x_{n-1}$. In this latter case, one can use iteratively the property in eq. \ref{prop} to solve these integrals up to the last one (in $x_0$), whose range is always $(a,b)$ and where $F(a)=0$, $F(b)=1$ can be finally applied, to give a result which will not depend on the precise shape of $f(x)$.\\

\noindent According to theorem 1, Gaussian, uniform, power law, etc, uncorrelated random series all have the same HVG motif profiles. As a byproduct, for any kind of sufficiently long time series $\{x_t\}_{t=1}^N$ where $x_t \in f(x)$ and $f(x)$ is continuous, if we randomize (shuffle) the time series, the motif profile of the randomized series is equal to eq. \ref{unif}. This is the reason why, at odds with the standard definition of a network's motif profile, for HVGs we don't need to rescale $\bf Z$ in any way to be able to compare across different time series and dynamical process.\\

\noindent Another notable consequence of theorem 1 is that it guarantees that series for which ${\bf Z}^4$ differ (even in the case of sufficiently long time series) from eq. \ref{unif} are not uncorrelated random series. This suggests a simple test for randomness \cite{pre}. For instance, one can use a Pearson's $\chi^2$ hypothesis test, where the null hypothesis is that the observed time series of $N$ data is random and uncorrelated (white noise). The test statistic is then
\begin{equation}
\chi^2 = (N-n)\sum_{i=1}^p \frac{[\mathbb{P}^4_i(\text{observed})-\mathbb{P}^4_i(\text{i.i.d.})]^2}{\mathbb{P}^4_i(\text{observed})}
\label{CHI}
\end{equation}
$\chi^2$ upper-critical values with $p-1$ degrees of freedom, for $p=6$ ($n=4$) are $11.07$ and $15.086$ at the $95\%$ and $99\%$ significance level (meaning that values of the $\chi^2$ larger than $11.07$ suggest that the observed series is not random at the $95\%$ significance level). More rigorously, as type-II motif is forbidden for aperiodic dynamics, we have only $p=5$ different motifs of size $n=4$, so the $\chi^2$ upper-critical values should be considered for 4 degrees of freedom: $9.49$ ($95\%$) and $13.28$ ($99\%$). 
%On the other hand,with 4 degrees of freedom, the probability that one misleadingly confounds the observed series with white noise is smaller than $5\%$ for $\chi^2<0.71$.

\subsection{Deterministic chaos: fully chaotic logistic map}
\begin{figure}
\includegraphics[width=9 cm]{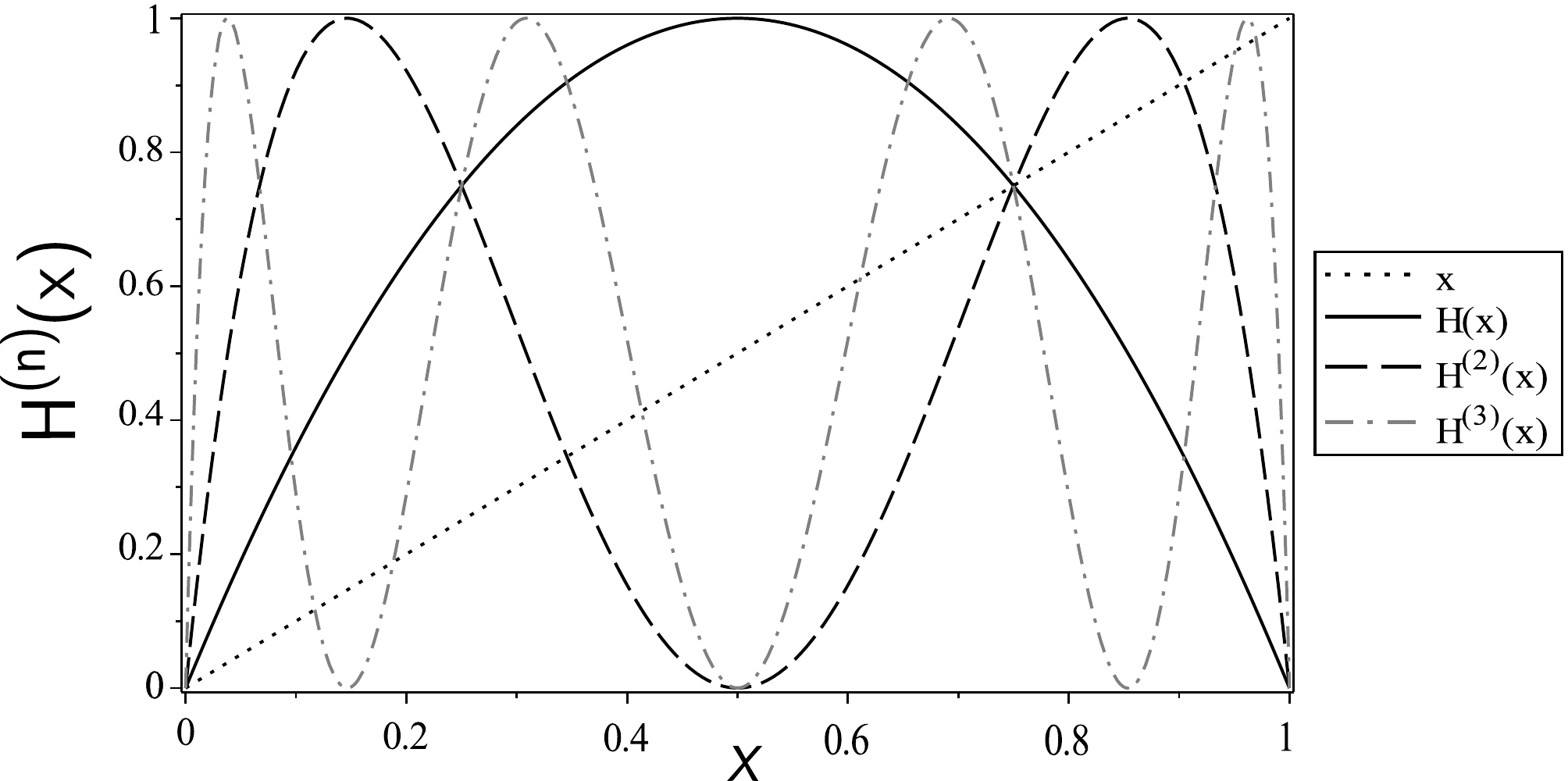}
\caption{Cobweb plot of the iterates of the fully chaotic logistic map ${\cal H}(x)=4x(1-x)$.}
\label{fig_label1}
\end{figure}

As previously stated, deterministic maps $x_{t+1}={\cal H}(x)$ are indeed Markovian, and for these situations the conditional probability is simply $f(x_2|x_1)=\delta(x_2-{\cal H}(x_1))$, where $\delta(x)$ is the Dirac-delta distribution. Therefore eqs. \ref{gen3}
 and \ref{gen4}, combined with inequality sets given in table \ref{ineq4} can be used to compute the motif profiles for different deterministic processes. In these cases, one  has to deal with simple integrals of the form
 
 \begin{equation}
\int_p^q \delta(x-y)dx=
\left\{
\begin{array}{rcl}
     1 & y \in [p,q]\\
     0 & \textrm{otherwise}
\end{array}
\right.
\label{theo}
\end{equation}
While in principle any deterministic process can be studied, we are interested in complex signals, so we focus on irregular, aperiodic dynamics. As a paradigmatic case, we tackle the fully chaotic logistic map $${\cal H}(x)=4x(1-x), \ x\in [0,1], \ f(x)=\frac{1}{\pi\sqrt{x(1-x)}}.$$ In this case, $f(x)$ is the invariant measure that describes in a probabilistic way the average time spent by a chaotic trajectory in each region of the attractor. Let us start by considering ${\bf Z}^3:=(\mathbb{P}^3_1,\mathbb{P}^3_2)$, for which
\begin{align}
&\mathbb{P}^3_1 = \nonumber\\
&\int_0^1 f(x_0)dx_0 \int_{x_0}^1 \delta (x_1 - {\cal H}(x_0)) dx_1 \int_0^1 \delta(x_2- {\cal H}^2(x_0) )dx_2, \nonumber \\
&\mathbb{P}^3_2 = \nonumber\\
&\int_0^1 f(x_0)dx_0 \int_{0}^{x_0} \delta (x_1 - {\cal H}(x_0)) dx_1 \int_{x_1}^1 \delta(x_2- {\cal H}^2(x_0) )dx_2.\nonumber
\end{align}
According to property in eq. \ref{theo}, the Dirac-delta integrals only have the effect of shrinking the range of integration of $x_0$. For instance, for $\mathbb{P}^3_1$, the integral in $x_1$ requires ${\cal H}(x_0) > x_0$, whereas the integral in $x_2$ simply requires ${\cal H}^2(x_0) \in [0,1]$. While the latter inequality is fulfilled for all $x_0 \in [0,1]$ (and thus has no effect), the former one requires $x_0 \in [0,3/4]$. This can be easily seen from the cobweb plot of ${\cal H}(x)$ and its iterates (see figure \ref{fig_label1}): ${\cal H}(x)>x$ for $x\in [0,3/4]$. Altogether, 
$$\mathbb{P}^3_1 = \int_0^{3/4} f(x_0)dx_0 = 2/3$$
On the other hand, motif normalization imposes $\mathbb{P}^3_2=1/3$. The same result is obviously found if we compute $\mathbb{P}^3_2$ explicitly: in this case the integral in $x_1$ requires ${\cal H}(x_0)<x_0$, which holds when $x_0 \in [3/4,1]$, and the integral in $x_2$ requires ${\cal H}^2(x_0)>x_1 \leftrightarrow {\cal H}^2(x_0)>{\cal H}(x_0)$. Looking at the cobweb plots, this final condition is met in two subintervals, so the intersection with the first condition yields a final interval $x_0 \in [3/4,1]$, for which 
$$\mathbb{P}^3_2 = \int_{3/4}^1 f(x_0)dx_0 = 1/3,$$
as expected. 
These results coincide with those found for i.i.d. series, meaning that ${\bf Z}^3$ doesn't capture enough structure to distinguish both processes. Let us proceed in an equivalent way to compute ${\bf Z}^4=(\mathbb{P}_1^4,\dots,\mathbb{P}_6^4)$. It becomes evident that integrals associated to $x_n$ deal with the cobweb plots of 
${\cal H}(x), {\cal H}^2(x), \dots, {\cal H}^n(x)$. Accordingly, these integrals are ultimately related with the structure of fixed points of ${\cal H}^n(x)$, and with the solutions of equations of the form ${\cal H}^r(x)={\cal H}^s(x)$ for some $r$ and $s$.
We only have algebraic closed expressions for the fixed points of ${\cal H}(x)\rightarrow \{0,3/4\}$ and ${\cal H}^2(x)\rightarrow \{0,\frac{5-\sqrt{5}}{8},3/4,\frac{5+\sqrt{5}}{8}\}$ (for $n\geq 3$, ${\cal H}^n(x)$ is a polynomial of order larger or equal to 6 and according to Abel-Ruffini's theorem, the set of fixed points does not have in general an algebraic expression, however we can compute them up to arbitrary precision). Other values of interest include the roots of ${\cal H}^3(x)={\cal H}^2(x)$, and specially the largest one $x = 1/2+\sqrt{3}/4$.\\
Let us show how to compute one of these motif probabilities. For instance,
\begin{widetext} 
\begin{align}
\mathbb{P}^4_5 = \int_0^1 f(x_0)dx_0 \int_0^{x_0} \delta (x_1 - {\cal H}(x_0)) dx_1 \int_{x_1}^{x_0} \delta(x_2- {\cal H}^2(x_0) )dx_2\int_{x_2}^1 \delta(x_3-{\cal H}^3(x_0) )dx_3 
\end{align}
\end{widetext} 
which reduces to 
$$\mathbb{P}^4_5 = \int_p^q f(x_0)dx_0,$$
where $[p,q]$ can be hierarchically obtained as:\\
${\cal H}(x_0)<x_0 \cap [0,1] \Rightarrow x_0 \in [3/4,1]$;\\
${\cal H}^2(x_0)<x_0 \cap {\cal H}^2(x_0)> {\cal H}(x_0) \cap [3/4,1]\Rightarrow x_0 \in [\frac{5+\sqrt{5}}{8},1] $;\\
${\cal H}^3(x_0)> {\cal H}^2(x_0) \cap [\frac{5+\sqrt{5}}{8},1] \Rightarrow x_0 \in [x_p,1]$, where $x_p$ is the second largest root fulfilling ${\cal H}^3(x_p)={\cal H}^2(x_p)$, i.e $x_p= 1/2+\sqrt{3}/4$. Altogether, 
\begin{align}
\mathbb{P}^4_5 =& \int_{1/2+\sqrt{3}/4}^1 \frac{1}{\pi\sqrt{x_0(1-x_0)}}dx_0= \nonumber \\
&\frac{1}{\pi} B_{\left[\frac{1}{2}+\frac{\sqrt{3}}{4},1\right]}\left(\frac{1}{2},\frac{1}{2}\right)=\frac{1}{6}(=4/24) \nonumber
\end{align}
(where $B$ is the incomplete Beta function), which is indeed quite different from the result found for i.i.d., $\mathbb{P}^4_5(\text{i.i.d.})=2/24$.\\
Similar arguments can be used to obtain analytically the rest of probabilities (explicit computations are put in an appendix), finding
\begin{eqnarray}
{\bf Z}^3=\bigg[\frac{2}{3},\frac{1}{3}\bigg]; \ {\bf Z}^4=\bigg[ \frac{8}{24},0,\frac{4}{24}, \frac{8}{24}, \frac{4}{24}, 0\bigg]
\label{log}
\end{eqnarray}

Comparing this set of motif probabilities with the result for i.i.d. (eq. \ref{unif}), we can conclude that ${\bf Z}^4$  distinguishes the fully chaotic logistic map from a purely uncorrelated stochastic process.  Note, of course, that a similar derivation can be performed in other deterministic maps; in this sense the methodology is general (however one encounters problems when the attractor has a fractal dimension, and one needs to carefully choose a proper integration theory). These exact results are also in excellent quantitative agreement with numerics performed in finite series (left panel of figure \ref{fig:1}), so convergence to the theory with series size is quite fast, enabling its use in empirical cases. To be more precise, in the next subsection we make a study of how fast results for short time series converge to the asymptotic theory as series size increases.

\subsection{Convergence of finite series}
\begin{figure}
\centering
\includegraphics[width= 8.5 cm]{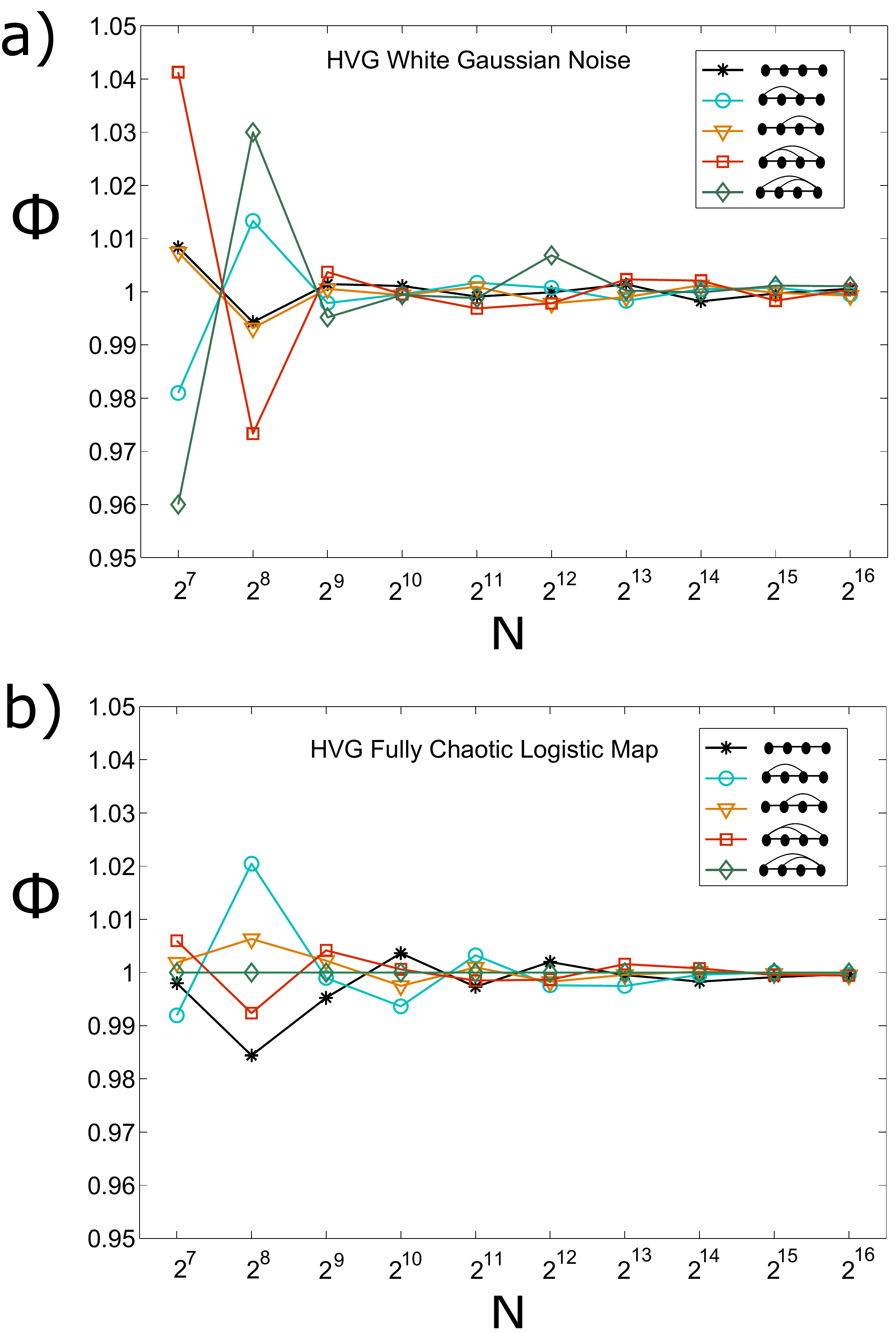}
\caption{(Color online) The measured frequency of appearance rescaled by its theoretical value $\Phi$ is plotted for each motif associated to Gaussian white noise (panel a) and to a fully chaotic logistic map (panel b) in function of the time series size $N$; results are averaged over 100 realisations. The curves oscillate with fast decreasing amplitude around the value 1 (for $2^{9}$ the amplitude is less than $10^{-2}$) indicating fast asymptotic convergence of the measured motif profile to the theoretical profile in both cases.}
\label{fig_robustness}
\end{figure}

In order to be more precise about the convergence speed of finite-size numerics to the theory (which in rigour only holds for bi-infinite time series), we have computed for series of size $N$ the numeral estimate ${\bf Z}^4(N)$ for both i.i.d. and the fully chaotic logistic map, and compare it with the asymptotic values ${\bf Z}^4$. Results are plotted in figure \ref{fig_robustness}, where we plot $\Phi(N)=\langle {\bf Z}^4 (N) \rangle / {\bf Z}^4$ as a function of the series size $N$ (the average is with respect to realizations). Results indicate that convergence to the asymptotic theory is already reached for $N\ll 10^4$ (which is the conservative size that is used all over this work).

\subsection{Stochastic processes with correlations}
\begin{figure}
\centering  
\includegraphics[width=8.5 cm]{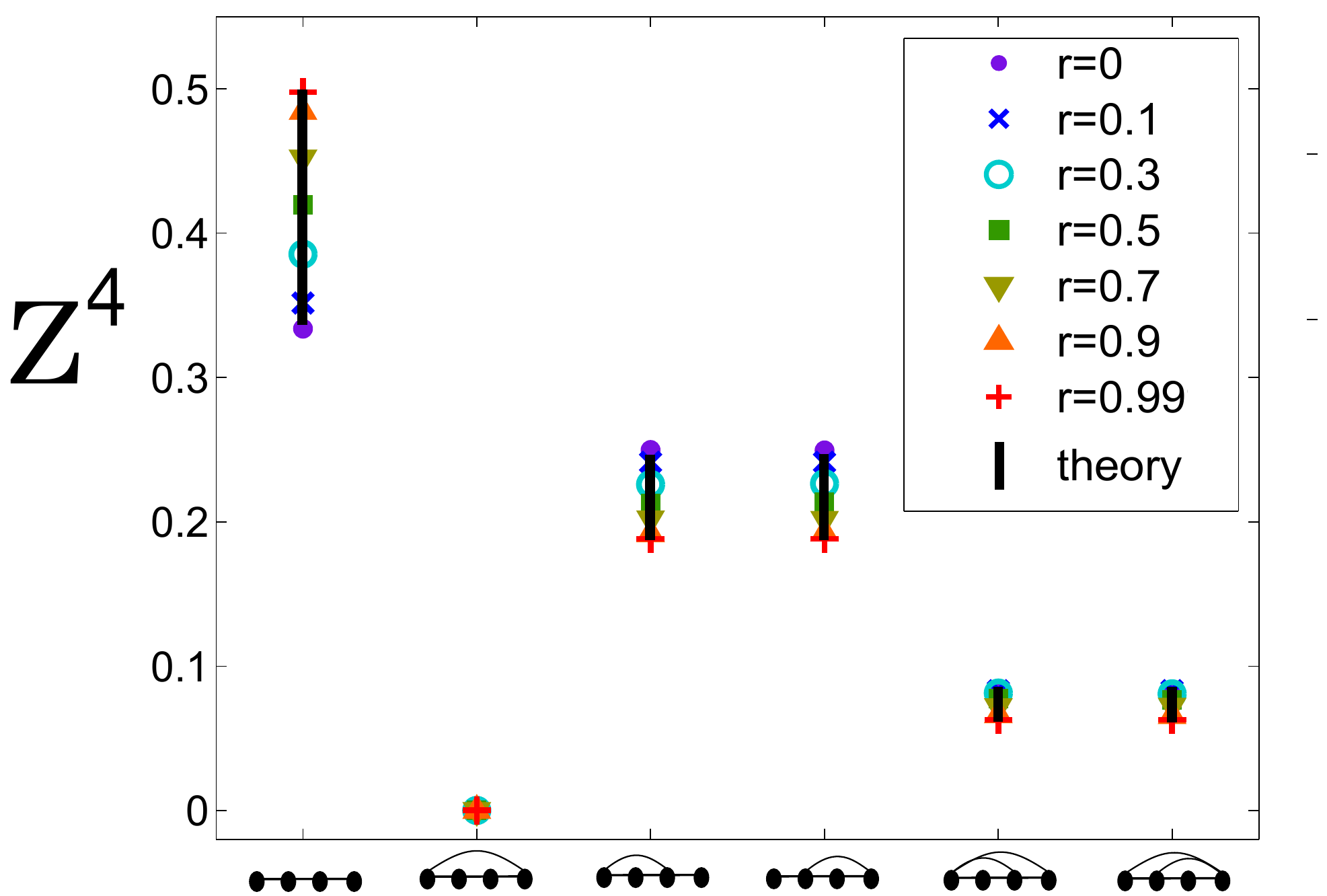}
\caption{(Color online) HVG significance profile ${\bf Z}^4$ for AR(1) processes described by eq.\ref{eq:rednoise}, for different values of the correlation coefficient $r$. When $r$ increases the appearance probability of motif of type-I increases while the rest of probabilities decrease. This is simply due to the fact that finding  constant sequences $x_{t+3}=x_{t+2}=x_{t+1}=x_{t}$ becomes more probable as $r$ increases.}
\label{fig:3}
\end{figure}

To round off the theory section, and to explore how results deviate from i.i.d. for correlated stochastic processes, we consider coloured noise with exponentially decaying correlations as described by the AR(1) process:
\begin{equation}    
\begin{cases}
x_0= \xi_0 \\
x_{t}= r x_{t-1}+\sqrt{(1-r^2)}\xi_{t}, \quad t\geq 1
\end{cases}    
\label{eq:rednoise}
\end{equation} 
where $\xi_{t} \sim {\cal N}(0,1)$ is Gaussian white, and $r$, $0<r< 1$ is a parameter that tunes the correlation. The auto-correlation function $C(t)$, which describes the correlation of the position at $x_{t_0}$ and $x_{t_0+t}$ decays exponentially $C(t)=e^{-t/\tau}$, where the characteristic time  $\tau=1/\ln(r)$. In the limit $r\to 0$, the correlations vanish and the process reduces to a white noise signal. The limit $r \to 1$ is more delicate, but intuitively in this limit the process gets completely correlated and tends to be constant $x_{t+1} = x_{t}$ $\forall t$.\\
This is a family of models parametrized by the coefficient $r$. For $0<r<1$, these models are indeed Gaussian, Markovian and stationary, with a probability density $f(x)$ and transition probability $f(x_2|x_1)$ are
\begin{equation}
\begin{cases}
f(x)=\frac{\exp(-x^2/2)}{\sqrt{2\pi}} \nonumber\\  
f(x_2|x_1)=\frac{\exp[-(x_2-rx_1)^2/(2(1-r^2))]}{\sqrt{2\pi (1-r^2)}}\nonumber
\end{cases}
\end{equation}
respectively. Since $x$ are Gaussian variables they can vary in $(-\infty,\infty)$. We focus on ${\bf Z}^4$ that we know gave good discriminatory results between i.i.d. and chaos. For illustration, the first element reads
\begin{widetext}
\begin{align}
\mathbb{P}^4_1= \int_{-\infty}^{\infty} \frac{e^{\frac{-x_0^2}{2}}}{\sqrt{2\pi}} dx_0 &\int_{-\infty}^{\infty} \frac{e^{\frac{-(x_1-rx_0)^2}{2(1-r^2)}}}{\sqrt{2\pi (1-r^2)}} dx_1 \int_{-\infty}^{x_1} \frac{e^{\frac{-(x_2-rx_1)^2}{2(1-r^2)}}}{\sqrt{2\pi (1-r^2)}} dx_2 \int_{-\infty}^{x_2} \frac{e^{\frac{-(x_3-rx_2)^2}{2(1-r^2)}}}{\sqrt{2\pi (1-r^2)}} dx_3 + \nonumber\\
&\int_{-\infty}^{\infty} \frac{e^{\frac{-x_0^2}{2}}}{\sqrt{2\pi}} dx_0 \int_{x_0}^{\infty} \frac{e^{\frac{-(x_1-rx_0)^2}{2(1-r^2)}}}{\sqrt{2\pi (1-r^2)}} dx_1 \int_{x_1}^b \frac{e^{\frac{-(x_2-rx_1)^2}{2(1-r^2)}}}{\sqrt{2\pi (1-r^2)}} dx_2 \int_{-\infty}^{\infty} \frac{e^{\frac{-(x_3-rx_2)^2}{2(1-r^2)}}}{\sqrt{2\pi (1-r^2)}} dx_3 
\label{}
\end{align}
\end{widetext}
For any particular value of $r$, these integrals can be evaluated up to arbitrary precision using Mathematica \cite{wolfram}. In table Table \ref{tablaAR} we report the theoretical values of ${\bf Z}^4(r)$ for $r\in [0.02 - 0.99]$. These are in perfect agreement with numerical simulations performed on finite series of size $N=10^4$ (ensemble averaged over 100 realizations) for $r=\{0,0.1,0.3,0.5,0.7,0.9,0.99\}$, as shown in Figure \ref{fig:3}. As $r>0$ the profiles deviate from i.i.d. and thus, again, these features can easily distinguish between exponentially coloured and white noise.

\begin{table}[]
\centering
\label{tab:numeric}
\begin{tabular}{l|l|l|l|l}
\bf r    &  \multicolumn{1}{l|}{$ \bf \mathbb{P}^4_1$} & \multicolumn{1}{l|}{$ \bf \mathbb{P}^4_2$}   & \multicolumn{1}{l|}{$ \bf \mathbb{P}^4_3$, $\mathbb{P}^4_4$} & \multicolumn{1}{l}{$ \bf \mathbb{P}^4_5$, $ \bf \mathbb{P}^4_6$}  \\ \hline 
0.02 & 0.3370                  & 0                       & 0.2482                  & 0.0833                  \\
0.04 & 0.3406                  & 0                       & 0.2464                  & 0.0833                  \\
0.06 & 0.3443                  & 0                       & 0.2446                  & 0.0832                  \\
0.08 & 0.3478                  & 0                       & 0.2429                  & 0.0831                  \\
0.1  & 0.3514                  & 0                       & 0.2412                  & 0.0830                  \\
0.2  & 0.3690                  & 0                       & 0.2333                  & 0.0822                  \\
0.3  & 0.3862                  & 0                       & 0.2260                  & 0.0809                  \\
0.4  & 0.4030                  & 0                       & 0.2192                  & 0.0793                  \\
0.5  & 0.4196                  & 0                       & 0.2130                  & 0.0772                  \\
0.6  & 0.4359                  & 0                       & 0.2072                  & 0.0748                  \\
0.7  & 0.4521                  & 0                       & 0.2018                  & 0.0722                  \\
0.8  & 0.4681                  & 0                       & 0.1967                  & 0.0692                  \\
0.9  & 0.4841                  & 0                       & 0.1920                  & 0.0660                  \\
0.95 & 0.4919                  & 0                       & 0.1897                  & 0.0643                  \\
0.97 & 0.4945                  & 0                       & 0.1888                  & 0.0636                  \\
0.99 & 0.4973                  & 0                       & 0.1879                  & 0.1879                 
\end{tabular}
\caption{Theoretical values of ${\bf Z}^4(r)$ for the AR(1) process evaluated at different values of the coefficient $r$.}
\label{tablaAR}
\end{table}

\section{Robustness}
In the preceding section we have developed a general theory to compute explicitly the motif profile of HVGs associated to a given type of dynamics. We have applied this theory to find theoretical expressions in the case of white and coloured noise as well as chaotic dynamics, and have shown that these predictions perfectly match the results found in numerical simulations for reasonably short time series. The theory (which is exact in the limit of infinite size series) is thus correct also in the case of short time series. These are nonetheless only idealized models: empirical time series, however, even if they comply to a particular dynamical system are usually polluted with measurement noise. Therefore, before being able to apply this new technique to real world phenomena, we need to assess its robustness and reliability against noise contamination. To do that, we consider a situation where a chaotic time series is contaminated with different amounts of white noise, and explore the ability of ${\bf Z}^4$ to detect the chaotic signal. Formally, we pollute a chaotic signal $x(t)$ with uniform white noise $\xi(a)$ and thus construct a noisy chaotic signal $Y(t)$ such that
\begin{equation}
\begin{cases}
Y_{t}= x_{t}+\xi\\
x_{t}= 4x_{t-1}(1-x_{t-1}) \\  
\xi \sim U[0,a],\quad 0\leq a\leq 1,
\end{cases}
\end{equation}
where $a$ tunes the noise power. The noise-to-signal ratio of the signal $Y_{t}$ is defined as $NSR=\sigma^2_{\xi}/\sigma^2_{Y}$ (where $\sigma^2_{\cdot}$ denotes the variance of signal $\cdot$), thus $NSR$ will increase monotonically with $a$. For $NSR\ll1$, the noise contamination is small. Any technique that is able to  distinguish $Y(t)$ and $\xi(t)$ for increasing values of $NSR$ is said to be robust to noise. For $NSR=1$ the levels of the signal and the noise contamination are comparable and for $NSR>1$ the underlying chaotic signal is effectively hidden. Of course, when $a$ reaches a certain value it won't be possible any more to distinguish the underlying chaotic nature of the time series by looking at the motif profile. To estimate this threshold we can use two different tests:\\

\begin{itemize}
\item The first test makes use of the ($L_1$) distance in motif space between the signal and the noise $d(a)= |{\bf Z}^4(Y)-{\bf Z}^4(\text{iid})|$. This is just a simple, motif-based similarity metric between two graphs, that we use here to measure the similarity between two series. Ideally, the threshold of distinguishability is the smallest value of $a$ for which $d(a)=0$. However, in practice, as we are dealing with finite size series, there will always be a small uncertainty associated to small finite-size deviations from the theory. That is, if one estimates the ${\bf Z}^4(\text{iid})$ with an ensemble average of $m$ realizations of a finite random time series of $N$ data, then for each element in the profile, the standard deviation of the estimate $\mathbb{P}^4_i$ will be a finite value (that converges to zero as $N$ and $m$ increases). We define $\sigma({\bf Z}^4(\text{iid}))$ as the vector where the $i$-th term is such standard deviation, for the same values of $N$ and $m$ used in the analysis of $Y(t)$. Then, we define the \textit{uncertainty threshold} $a^*$ as the smallest value of $a$ such that $d(a)\leq |\sigma({\bf Z}^4(\text{iid}))|$ (intuitively, $a^*$ is the smallest value for which we don't know if the difference in the motif profile between the empirical results and the theory are due to the fact that there is a chaotic signal underlying the process, or just due to finite size effects).
\item The second possibility is to use a Pearson's $\chi^2$ hypothesis test such as equation \ref{CHI} with 4 degrees of freedom, where the null hypothesis is that $Y(t)$ (the observed series) is just white noise (no hidden signal). 
In this latter case, we are not taking into account the deviations associated to finite size effects in the profile of i.i.d., though. If $\chi^2<9.49$, then we can't reject the null hypothesis at the $95\%$ significance level: this is the limit of what we could call certain distinguishability. 
%However this is a very conservative way of accepting the null hypothesis. When
% $\chi^2<0.71$, this means that the amount of noise that have been added to the chaotic signal is such that the probability of the null hypothesis to be false is at most $5\%$. This is the limit for certain undistinguishability. According to this test, $a^*$ lies somewhere between certain distinguishability and uncertain undistinguishability.
\end{itemize}

\begin{figure*}
\centering 
\includegraphics[width= 18 cm]{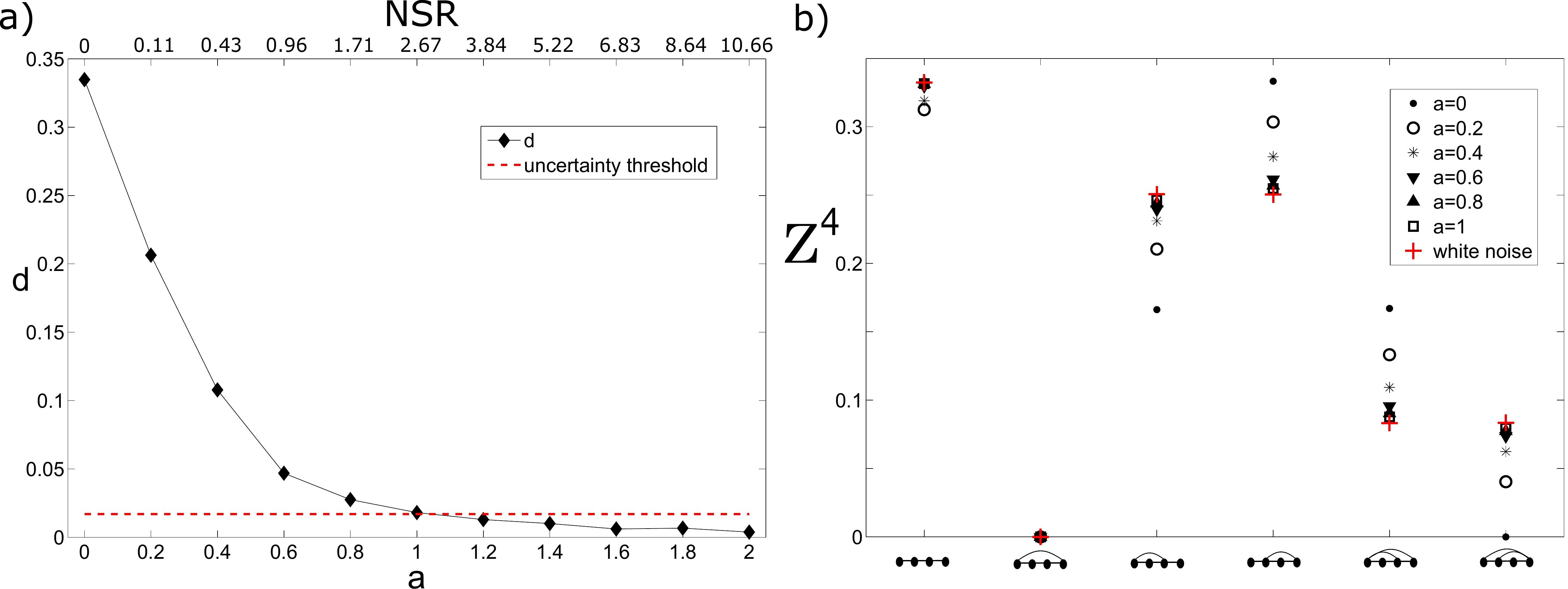}
\caption{(Color online) Robustness of motif profiles for chaotic series (fully chaotic logistic map) polluted with white noise. Panel a): by increasing the amount of extrinsic noise (parameterised by $a$) the distance in motif space between the noisy chaotic signal and white noise decreases (see the text). The method is extremely robust as one can distinguish the noisy chaotic signal from pure white noise up to a noise-to-signal ratio $NSR\approx 2.67$. Panel b): the 4-node motif profile of the noisy chaotic signal $Y_{t}$ for different degrees of noise contamination ($a$). Motifs III, IV, V and VI are the most informative as they concentrate most of the profile variability.}
\label{fig:2}
\end{figure*}

\noindent For each value of the parameter $a$, we have simulated a time series of $N=10^4$ steps from the process $Y(t)$, and results were ensemble averaged over $m=100$ realisations. In panel (b) of figure \ref{fig:2}, we plot the motif profile as a function of $a$. It is interesting to observe that the probabilities which vary most with $a$ are related to types III, IV, V and VI, while type-I seems to maintain approximately the same rate of appearance (we will show later that this is not always the case). In the panel (a) of the same figure we plot $d(a)$.  As expected, $d(a)$ is a monotonically decreasing function of $a$, and we find  $a^*\approx1$. Remarkably, this corresponds to a value of the noise to signal ratio $NSR\approx 2.67$. This is indeed confirmed by the Pearson $\chi^2$ test, where we found that the limit for confidently rejecting the null hypothesis -certain distinguishability- is $a\approx 1$ (i.e. $NSR\approx 2.67$). 
%Now, in order to confidently accept the null hypothesis -certain undistinguishability- one needs contaminate the signal with noise up to  $a\approx 2$ (i.e. $NSR\approx 10.66$). 
These results prove that ${\bf Z}^4$ is indeed an extremely robust feature with respect to measurement noise contamination, hence useful for applications.

\section{Principal component analysis}
\label{sec:automatic}

According to the last sections, we can conclude that the HVG ${\bf Z}^4$ is an informative feature of complex dynamics. Here we summarise and gather the findings on i.i.d., fully chaotic logistic maps (with and without noise contamination) and coloured noise, and we complement those with additional chaotic maps (Ricker's map, Cubic map, Sine map). Each process is described by the six dimensional vector ${\bf Z}^4$ (although in practice this space is 5-dimensional as $\mathbb{P}^4_2=0$). As this representation is obviously not very convenient for readability, we have projected each point into a 2-dimensional space spanned by the principal components of the data. We recall that Principal Component Analysis (PCA) \cite{PCA} is a common statistical procedure to perform dimensionality reduction on data. It uses an orthogonal transformation to project our set of observations, originally described in $\mathbb{R}^{6}$ -where each direction describes the probability of occurrence of a given motif, this being possibly correlated among observations- into a lower dimensional subspace spanned by the so called principal components, obtained from the eigenvectors of the dataset covariance matrix. These particular directions are such that (i) they are orthogonal, (ii) the first principal component has the largest possible variance (that is, accounts for as much of the variability in the data as possible), and each succeeding component in turn has the highest variance possible under the constraint that it is orthogonal to (i.e., uncorrelated with) the preceding components. 
%Thus
%projecting each observation $O_i$ (originally $O_i \in \mathbb{R}^{6}$) into a %smaller space spanned by the first $m$ principal components hugely reduces the %dimensionality of the observations, while keeping the relevant information of the %data. This projection is indeed the one that minimizes the mean squared distance between the data points and their projections. 
If the data can be efficiently projected in a lower dimensional space, then the eigenvalues associated to each of the principal components sum up a large percentage of the data variability. In that case, the projection is said to be faithful, and constitutes an accurate description of the data.\\
To summarise, the following processes have been considered (for all of them, we have estimated $\bf{Z^4}$ from a time series of $N=10^4$ points, and have averaged this over 100 realisations):\\
\begin{itemize}
\item White noise (i.i.d.) with Gaussian, exponential, uniform and power-low probability densities.% All these processes share the same ${\bf Z}^4$ according to the theory, so all these processes have the same coordinates in the original six-dimensional space. 
\item Chaotic maps, in particular: Fully chaotic logistic map $x_{t+1}= 4x_{t}(1-x_{t})$, 
Ricker's map $x_{t+1}= 20x_{t}e^{-x_{t}}$, 
Cubic map $x_{t+1}= 3x_{t}(1-x_{t}^2)$, Sine map $x_{t+1}= \sin(\pi x_{t})$
\item Noisy logistic map with $a=\{0.2,0.4,0.6,0.8,1.0\}$
\item Coloured noise for $r=\{0.1,0.3,0.5,0.7,0.9,0.99\}$
\end{itemize}
   
\begin{figure*}
\centering
\includegraphics[width=13 cm]{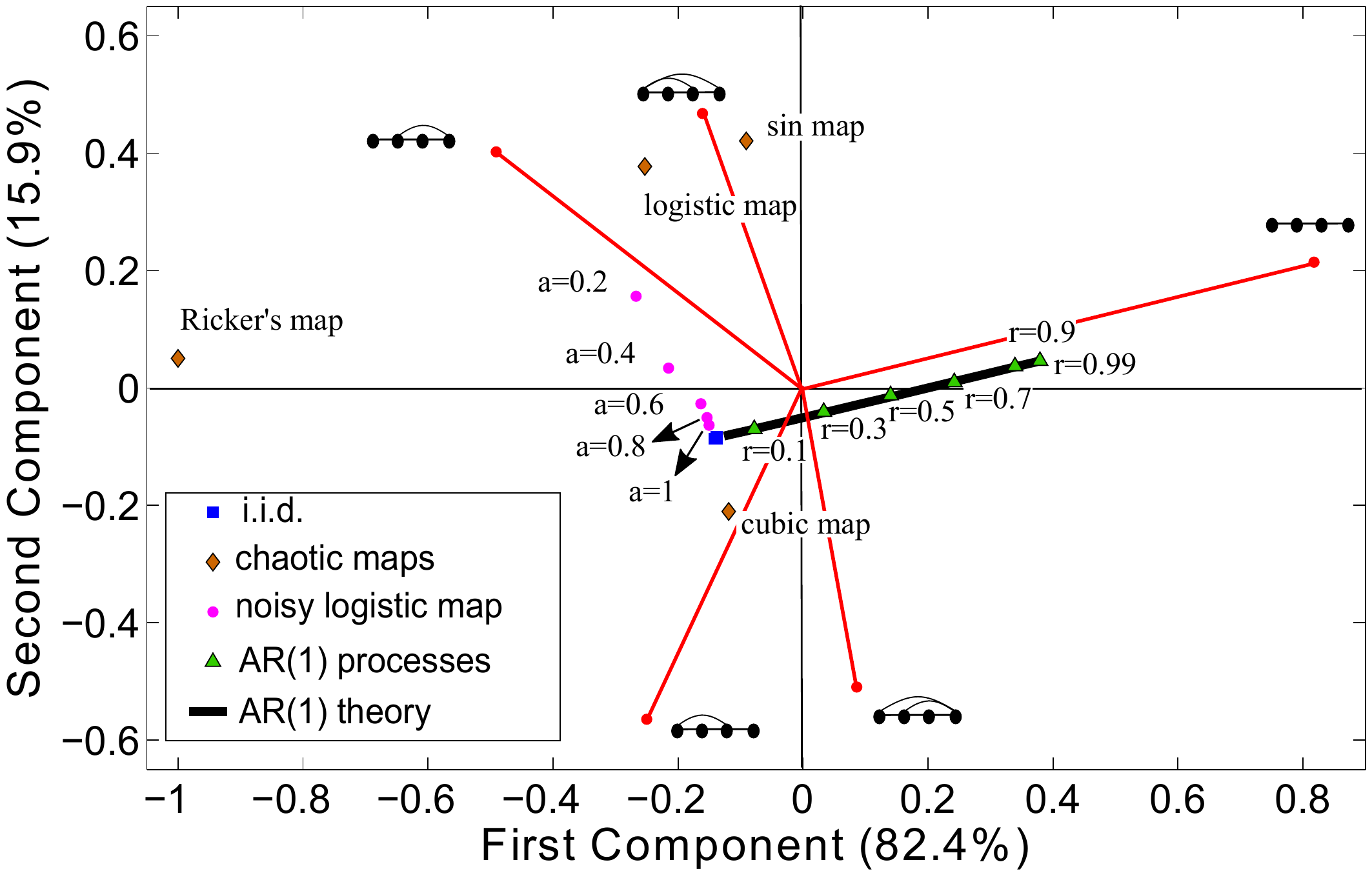}
\caption{(Color online) 2-dimensional projection obtained via Principal Component Analysis on ${\bf Z}^4$ for time series generated from different deterministic and stochastic processes: different white noise series respectively with Gaussian, exponential, uniform and power low (blue squares), chaotic maps (brown diamonds), noisy logistic map for different levels of contamination (purple dots) and different stochastic correlated AR(1) processes (green triangles). The relative weight of each motif in this projection principal components is also plotted using red solid axes.}   
\label{fig:4}
\end{figure*}
The projection into the space spanned by the first two principal components is shown in figure \ref{fig:4}. Interestingly, these first two components capture about $98.3\%$ of the variability of the set of variables $\{{\bf Z}^4\}$. This means that motif probabilities are indeed highly correlated, and as few as two real  numbers per time series seem already enough to describe them. 
%As expected, white noise is well distinguished from all the chaotic maps points. Coloured noise, which interpolates between white noise ($r\to 0$) and a constant series ($r \to 1$) is projected into a straight line that departs from i.i.d. as $r$ increases in the direction where type-I motif increases, as it should. In another direction, the noisy logistic map deviates from a purely chaotic logistic map as $a>0$ and approaches i.i.d. as the noise level increases. We also plot, as red solid axes, the projection of each motif in this new basis (see also table \ref{tablemot}).\\  
The patterns related to the different processes in this 2-dimensional component space help visualize some of the results previously found and make interesting considerations: 
\begin{itemize}
\item  All the i.i.d. processes have the same coordinates in the 2-dimensional space which do not correspond to the coordinates of any other class of processes considered. Indeed according to the theory, i.i.d. processes share the same ${\bf Z}^4$. 
\item Red solid axes (color online) describe the projection of each motif in this new basis (see also table \ref{tablemot}) and give an idea of which motif types are more related to different processes, thus helping to interpret a particular trajectory in this space, as a given process changes. For instance 
coloured noise which interpolates between white noise ($r\to 0$) and a constant series ($r \to 1$) projects into a straight line-like trajectory, departing from the i.i.d. coordinates and following the direction where type-I motif increases as $r$ increases. Analogously, as the noise level $a$ increases the noisy logistic map interpolates between the fully chaotic logistic coordinate and i.i.d. following a specific path.
\item The distance in this space between i.i.d. and the $(a=1)$-noisy logistic map gives us a rough idea of the distinguishability or coarse-graining distance, a lower bound below which any two processes cannot be distinguished.
\end{itemize}  
We conclude that ${\bf Z}^4$ is a highly informative and robust feature, which in principle could be used to assess similarities and differences across empirical complex signals. To test this hypothesis, in the final section we will explore this idea and will show that clustering of complex physiological processes is possible with this simple feature.

\begin{table}[]
\centering
\label{tab:3}
\begin{tabular}{l|l|l}
  &First Component & Second Component\\
  \hline
\begin{minipage}{.1\textwidth}
\includegraphics[width= 0.7\textwidth]{4motif-1.pdf}
\end{minipage} & 0.814       & 0.2114           \\
\begin{minipage}{.1\textwidth}
\includegraphics[width= 0.7\textwidth]{4motif-2.pdf}
\end{minipage} & -8.6e$^{-18}$       & 1.5e$^{-16}$           \\
\begin{minipage}{.1\textwidth}
\includegraphics[width= 0.7\textwidth]{4motif-3.pdf}
\end{minipage} & -0.2506       & -0.5670           \\
\begin{minipage}{.1\textwidth}
\includegraphics[width= 0.7\textwidth]{4motif-4.pdf}
\end{minipage} & -0.4926           & 0.4030           \\
\begin{minipage}{.1\textwidth}
\includegraphics[width= 0.7\textwidth]{4motif-5.pdf}
\end{minipage} & -0.1569       & 0.4608       \\
\begin{minipage}{.1\textwidth}
\includegraphics[width= 0.7\textwidth]{4motif-6.pdf}
\end{minipage} & 0.0852       & -0.5090           
\end{tabular}
\caption{Weights of each motif in the 2-dimensional projection of the set of all dynamical processes analysed (i.i.d. white noise, coloured noise with exponentially decaying correlations, chaotic maps, noisy chaotic logistic map).}
\label{tablemot}
\end{table}

\section{Unsupervised learning: disentangling meditative from other relaxation states using HVG motif profiles from heart rate time series}

It is well-known that meditation has a measurable effect on well-being. In particular, neuroscience has shown that meditation promotes EEG high-amplitude gamma synchronisation \cite{meditationPNAS}, or increases sustained attention \cite{meditation2} among others effects on the brain \cite{medi}. In this final section we explore, via a HVG motif profile analysis, if one can distinguish purely meditative states from general states of relaxation by only looking at a single physiological indicator: the heart rate series \cite{meditation_VG, meditation_fractals}. This analysis is based on experiments performed in a former publication \cite{peng1999exaggerated}. Data are freely available online \cite{physionet}.

\subsection{Data}
\begin{figure}
\centering
\includegraphics[width= 9 cm]{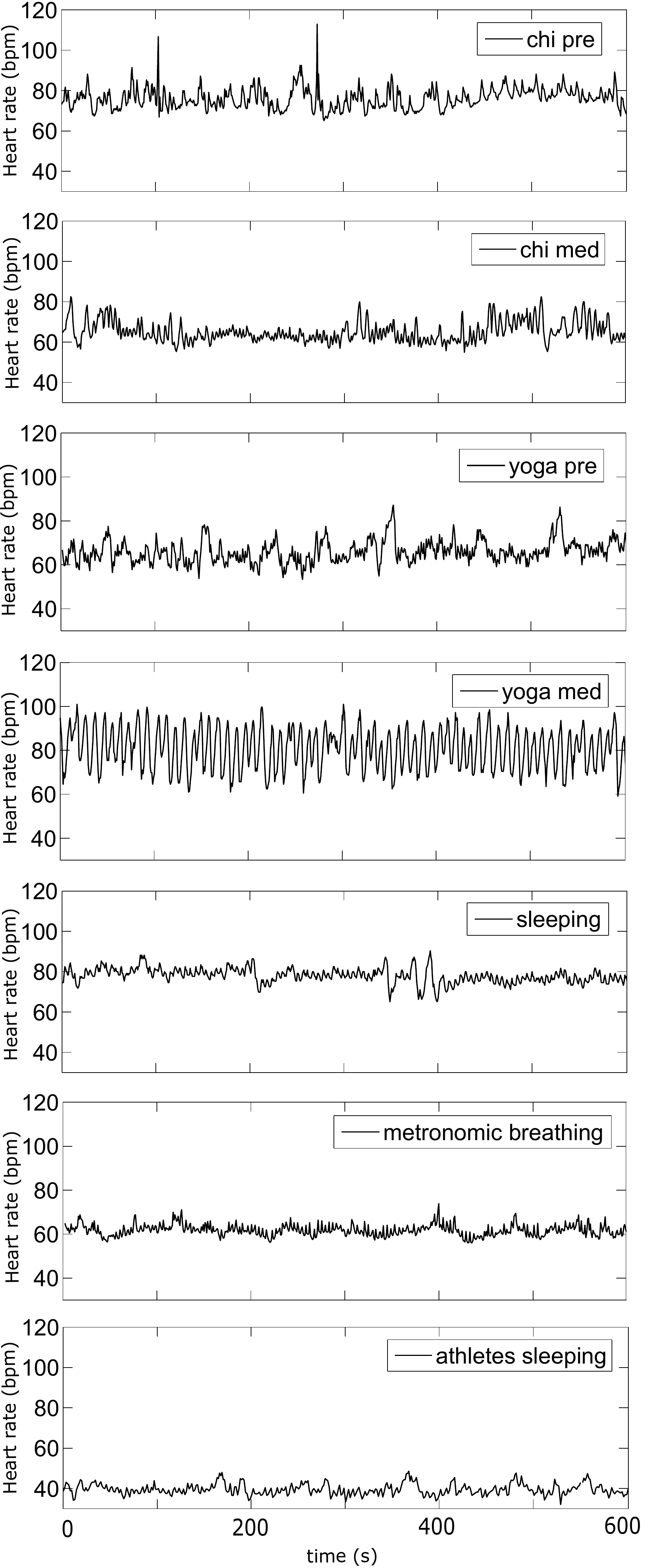}
\caption{Sample heart rate time series from patients in meditative and non-meditative states.}
\label{series_2}
\end{figure}

 Data are collected for five different groups of healthy subjects \cite{peng1999exaggerated}:
 \begin{itemize}
 \item The first group of 4 subjects (two women and two men in the age range 20-52) were \textit{expert} Kundalini Yoga meditators. Their heart rate was recorded for approximately fifteen minutes before the Yoga practice (pre-meditative state) and for approximately one hour during the breathing and chanting exercises (meditative state) (a total of 8 time series);
 \item The second group comprised 8 Chinese Chi Meditation practitioners, (five women and three man in the age range 26-35) \textit{relatively novice} in the practice. The heart rate of the subjects was recorded for approximately five hours during the pre-meditation (pre-meditative state) and for approximately one hour during the meditation session (meditative state)(a total of 16 time series).
 %, when the meditators sat quietly and were instructed by the master to breathe spontaneously while visualizing the opening and closing of a perfect lotus in the stomach (a total of 16 time series);
 \end{itemize}
 To better compare the pre-meditation and meditation states, three healthy, non-meditating control groups were considered from a database of retrospective electrocardiogram (ECG) signals:
 \begin{itemize}
 \item a spontaneous breathing group of 13 subjects (eight women and five men in the age range 25-35) during sleeping hours (general relaxation state) (a total of 13 time series);
 \item a group of 9 elite triathlon athletes (six women three men, age range 21-55) in the pre-race period during sleeping hours (general relaxation state) (a total of 9 time series);
 \item a group of 14 subjects (nine women and five men, age range 20-35) during supine metronomic breathing at 0.25 Hz (a total of 14 time series);
 \end{itemize} 
Sample time series from each group are plotted in figure \ref{series_2}.
In the original study the authors addressed the frequency spectra and observed prominent heart rate oscillations in the time series recorded during the two meditation practices with a peak in the range 0.025-0.35 Hz, and an overall variability of these series with respect to those from non-meditative states.  

\subsection{Unsupervised clustering based on HVG motif profiles}
The total dataset is made of a total of 60 time series (60 observations). A priori, we assume that each series is a different process. For each subject and state, we extract from the heart beat series the corresponding ${\bf Z}^4$ (detailed results are put in an appendix).\\

\begin{figure*}
\centering
\includegraphics[width=16 cm]{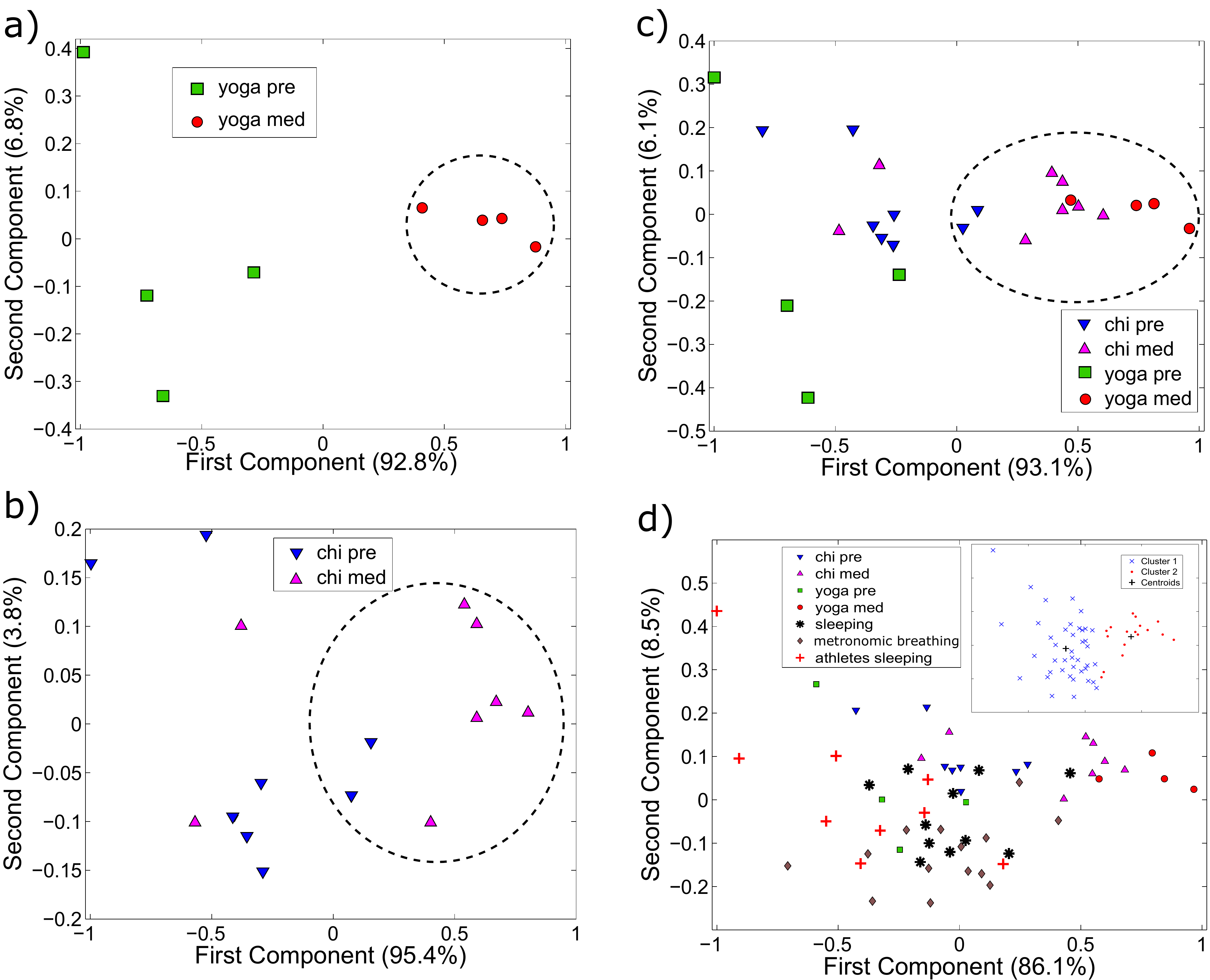}
\caption{(Color online) 2-dimensional Principal Component space of ${\bf Z}^4$ extracted from heart rate time series of subjects performing different tasks. (a) 4 Yoga meditators recorded during meditation (red dots) and during pre-meditation (green squares). The $k$-means algorithm (black dotted line) correctly assigns each of the 8 observations into the correct cluster. (b) 8 Chi meditators recorded during meditation (magenta triangles) and during pre-meditation (blue reverse triangles). The $k$-means algorithm correctly 12 out of 16 observations, however in this case subjects were novice meditators, hence clusters are not that well defined (see the text). (c) The two clusters found by the $k$-means algorithm correctly clusters the points related to Yoga meditation and Yoga pre-meditation, and 12 out of 16 points related to different Chi meditators (black dotted line). (d): although the $k$-means clustering (small panel on the top right) fails to precisely distinguish a cluster related to meditation from a cluster related to non-meditation, all the meditation points are surprisingly well separated from the all remaining points, on the right side of the plane.}   
\label{fig:6}
\end{figure*}

\noindent As a first analysis, we only consider the \textit{expert} meditators (first group) performing two different tasks and we explore if ${\bf Z}^4$ can disentangle the two tasks. Results are shown in panel (a) of figure \ref{fig:6}. In PCA space, we have 8 points scattered over the subspace spanned by the first two principal components. These aggregate more than $99\%$ of the data variance and is thus a faithful projection. Interestingly, already a visual inspection clusters the 4 subjects in the meditative state (red circles, right hand side of the plane) from those in the pre-meditative state (green squares). A simple $k$-means algorithm \cite{PCA} with $k=2$ correctly distinguishes the two states by assigning different clusters to both states (a black dotted oval is depicted with the purpose of visualizing the result of the $k$-means clustering).\\

\noindent In a second step, we consider the second group, formed now by {\it novice} Chi meditators before and during the practice. We repeat the analysis in the panel (b) of figure \ref{fig:6}. Again the first two principal components capture more than $99\%$ of the variability of the motifs considered. The scores related to the first principal component are very close to the ones found for the Yoga data subset (see appendix). For this meditation technique however it is not that easy to perfectly distinguish pre-meditative from meditative state clusters: the partition obtained with the $k$-means algorithm with input $k=2$ (visualized by the black dotted line) contain `false meditators' and `false non-meditators'. In order to quantify the performance of the clustering we use the so called \textit{purity coefficient} \cite{purity} defined by:
\begin{equation}
\text{purity}= \frac{n^1_m + n^0_n}{n^1_m + n^0_m + n^0_m + n^1_n}
\end{equation}
where $n^1_m$ is the number of meditators in the cluster 1, which is defined as the cluster where most of the meditators are found; $n^0_m$ is the number of meditators in the cluster 0, which is defined as the cluster where most of the non-meditators are found (`false non-meditators'); $n^1_n$ is the number of non-meditators in the cluster 1 (`false meditators'); $n^0_n$ is the number of non-meditators in the cluster 0.
Purity takes value in $[0,1]$ and was measured for the different partitions reported in figure \ref{fig:6} (see table \ref{puritable}); in this case we found $\text{purity}\simeq0.83$.
%Now, as in this experiment the subjects were inexperienced Chi meditators, it is plausible that some of them were not able to concentrate of perform the task adequately, what would put their motif profile mixed amongst the pre-meditative state subjects. \textcolor{red}{If this speculation} were to be the case, then it should be much more likely to find `false non-meditators' (i.e. subjects which were supposed to be in a meditative state but couldn't reach the adequate level of concentration and effectively were in a non-meditative state) than `false meditators'.
Now, as in this experiment the subjects were inexperienced Chi meditators, it is plausible that some of them were not able to concentrate of perform the task adequately, what would put their motif profile mixed amongst the pre-meditative state subjects.
As we can see in the figure, there is some evidence of finding the `false non-meditators' intertwined among non-meditators, but not the `false meditators' intertwined among meditators.\\

\noindent We then perform the same analysis by considering data from the first two groups (Yoga group and the Chi group) altogether. Here we also aim at distinguishing meditative from pre-meditative states, however this is in principle much more delicate and problematic as we have different subjects performing different tasks. The results are reported in panel (c) of Figure \ref{fig:6}, and are consistent with the first two analysis conducted before. In PCA space, the first two principal components still capture more than $99\%$ of the data variability (scores are reported in the appendix). $k$-means correctly clusters together most of the pre-meditative states and distinguishes them from the meditative states (Yoga and Chi-style), with purity$=0.75$. There are two clear 'false non-meditators' which seem to correspond to two novice Chi meditators that falsely fall in the non-meditation state despite they were supposedly performing meditation. The two `false meditators' are not mixed among the meditators but placed in the boundary of the cluster, meaning that a refined clustering algorithm would very likely do a better job. On the other hand, it is worth highlighting that meditators show lower scattering than non-meditators, and are placed at the right hand side of the plane. Among these, Chi meditators (the experienced subjects) appear even more towards the right hand side in the PCA plane. According to the motif scores (appendix), one can conclude that meditation promotes the onset of type-I motifs, that is to say, generates a relative decrease of high-frequency heart rate fluctuations.\\

\noindent Finally, in panel (d) of Figure \ref{fig:6} we show the results for the analysis of the whole data set (the projection in PCA space still gathers more than $94\%$ of the data variability). Here we have highly heterogeneous subjects performing totally different tasks, which somehow can be classified into `meditative' and `non-meditative' states. In the inset panel of the same figure, each observation is labelled according to the result of k-means (crosses for non-meditative and dots for meditative states). Despite the heterogeneity of subjects, the purity of the partition obtained is high ($\simeq0.81$), and most of the observations associated to the meditative state concentrate towards the right hand side of the PCA plane (which, again according to the scores, corresponds to an overcontribution of type-I motif).  
We conclude that meditative practices leave a unique physiological fingerprint in the heart rate time series of its practitioners, which can be distinguished from other relaxation techniques and states such as metronomic breathing or sleeping by using the HVG motif profile of each time series. This is a remarkable result, taking into account that this profile only consists of a vector of 6 numbers (actually 5 as $\mathbb{P}^4_2=0$) per observation.

\begin{table}[]
\centering

\begin{tabular}{|l|l|l|l|}
\hline
        Panel a & Panel b  & Panel c & Panel d \\ \hline
 1    & 0.83 & 0.75      & 0.81    \\
 \hline  
\end{tabular}
\caption{Purity measures \cite{purity} of the k-means clustering analysis depicted in the four panels of figure \ref{fig:6}: yoga meditators (panel a), chi meditators (panel b), yoga and chi meditators (panel c) and all states (panel d).}
\label{puritable}
\end{table}

\section{Conclusions}
The theory of visibility graphs (VG and HVG) allows us to describe and characterise time series and dynamics using the powerful machinery of graph theory and network science. Here we have introduced the concept of Horizontal Visibility Graph (HVG) motifs, substructures present in the HVG of a time series, whose statistics have been shown to be informative about the time series structure and its underlying dynamics (comparison with VG motifs will be published elsewhere \cite{inprep}). We have advanced a mathematically sound theory by which the motif profile of large classes of stochastic and deterministic dynamics can be computed exactly. Interestingly, under the HVG framework, graph motifs are in direct correspondence with ordinal patterns \cite{PE1,PE2,PE3,PEbook}. This means, for instance, that the theory developed here can be exported to find rigorous results on the permutation entropy \cite{PE2} and permutation spectra \cite{PE3} of different dynamical systems. In the same vein, one could import concepts and ideas from ordinal patters to the context of visibility graphs. For instance, one can define an HVG motif entropy $S_n = -\frac{1}{n}\sum \textbf{Z}_i^n \log (\textbf{Z}_i^n)$ and explore its similarities with permutation entropy. More generally, the relation (and possible equivalences) between ordinal pattern analysis (so called permutation complexity \cite{PEbook, PE2}) and horizontal visibility graph analysis should be studied in more depth.
We have found that this graph feature is surprisingly robust, in the sense that it is still able to distinguish amongst different dynamics even when the signals are polluted with large amounts of measurement noise, what enables its use in practical problems. Despite the apparently difficult combinatorial interpretation of the visibility criteria, these latter results further suggest that HVG motifs are more than just an arbitrary partition on the set of ordinal patterns.
As an application, we have tackled the problem of disentangling meditative from general relaxation states from the HVG motif profiles of heartbeat time series of different subjects performing different tasks. We have been able to provide a positive, unsupervised solution to this question by applying standard clustering algorithms on this simple feature.\\

To conclude, HVG motifs provide a mathematically sound, computationally efficient and highly informative simple feature (a few numbers per time series) which can be extracted from any kind of time series and used to describe complex signals and dynamics from a new viewpoint. In direct analogy with the role played by standard motifs in biological networks, further work should evaluate whether HVG graph motifs can be seen as the building blocks of time series. In this sense a study of standard network motifs on visibility graphs can be of interest, especially in the case of directed and weighted HVG where the edge weights describe temporal relations between nodes. 
Potential applications of visibility graph analysis pervades the biological, financial and physical sciences. 
 Finally, other questions for future work include to assess which motifs are more informative for a given class of dynamics, and to extend this analysis to the realm of multivariate time series \cite{multivariate}.

\newpage

\section*{APPENDIX I: explicit computation of ${\bf Z}^4$ for the fully chaotic logistic map}
\begin{itemize}

\item $\mathbb{P}^4_1$
\begin{widetext}
\begin{align}
\mathbb{P}^4_1 = \int_0^1 f(x_0)dx_0 &\int_0^1 \delta (x_1 - {\cal H}(x_0)) dx_1 \int_0^{x_1} \delta(x_2- {\cal H}^2(x_0) )dx_2\int_{0}^{x_2} \delta(x_3-{\cal H}^3(x_0) )dx_3 + \nonumber \\
&\int_0^1 f(x_0)dx_0 \int_{x_0}^1 \delta (x_1 - {\cal H}(x_0)) dx_1 \int_{x_1}^1 \delta(x_2- {\cal H}^2(x_0) )dx_2\int_0^1 \delta(x_3-{\cal H}^3(x_0) )dx_3 \nonumber
\end{align}
\end{widetext}
the first integral on the right gives the following conditions:\\
${\cal H}^3(x_0)< {\cal H}^2(x_0)$\\
${\cal H}^2(x_0)< {\cal H}(x_0)$\\
which are never satisfied. The second integral gives:\\
${\cal H}^2(x_0)> {\cal H}(x_0)$\\
${\cal H}(x_0)>x_0$\\
which are satisfied for $x_0 \in [0,1/4]$. Thus
 
$$\mathbb{P}^4_1 =\frac{1}{\pi} B_{\left[0,\frac{1}{4}\right]}\left(\frac{1}{2},\frac{1}{2}\right)=\frac{1}{3}\quad(=8/24).$$

\item 
$\mathbb{P}^4_2=0$ since the probability of having ${\cal H}^2(x_0) = {\cal H}(x_0)$ is of zero measure.\\

\item $\mathbb{P}^4_3$ 
\begin{widetext}
\begin{align}
\mathbb{P}^4_3 = \int_0^1 f(x_0)dx_0 &\int_{0}^{x_0} \delta (x_1 - {\cal H}(x_0)) dx_1 \int_{x_1}^{x_0} \delta(x_2- {\cal H}^2(x_0) )dx_2\int_{0}^{x_2} \delta(x_3-{\cal H}^3(x_0) )dx_3 + \nonumber \\
&\int_0^1 f(x_0)dx_0 \int_0^{x_0} \delta (x_1 - {\cal H}(x_0)) dx_1 \int_{x_0}^{1} \delta(x_2- {\cal H}^2(x_0) )dx_2\int_{0}^{1} \delta(x_3-{\cal H}^3(x_0) )dx_3 \nonumber
\end{align}
\end{widetext}
In the first term:\\
${\cal H}(x_0)<x_0 \Rightarrow x_0 \in [3/4,1]$\\
${\cal H}^2(x_0)> {\cal H}(x_0) \cap {\cal H}^2(x_0)<x_0 \cap [3/4,1]\Rightarrow x_0 \in [\frac{5+\sqrt{5}}{8},1]$\\
${\cal H}^3(x_0)< {\cal H}^2(x_0) \cap [\frac{5+\sqrt{5}}{8},1] \Rightarrow x_0 \in [\frac{5+\sqrt{5}}{8}, \frac{1}{2} + \frac{\sqrt{3}}{4}]$
Analogously for the second term,\\
${\cal H}(x_0)<x_0 \Rightarrow x_0 \in [3/4,1]$\\
${\cal H}^2(x_0)>x_0 \cap [3/4,1] \Rightarrow x_0 \in [3/4, \frac{5+\sqrt{5}}{8}]$\\
Altogether, 
$$\mathbb{P}^4_3= \frac{1}{\pi}B_{[3/4, \frac{1}{2} + \frac{\sqrt{3}}{4} ]}(1/2,1/2)=\frac{1}{6}(=4/24)$$

\item $\mathbb{P}^4_4$
\begin{widetext}
\begin{align}
\mathbb{P}^4_4 = \int_0^1 f(x_0)dx_0 &\int_{x_0}^1 \delta (x_1 - {\cal H}(x_0)) dx_1 \int_0^{x_1} \delta(x_2- {\cal H}^2(x_0) )dx_2\int_{x_2}^1 \delta(x_3-{\cal H}^3(x_0) )dx_3 + \nonumber \\
&\int_0^1 f(x_0)dx_0 \int_0^{x_0} \delta (x_1 - {\cal H}(x_0)) dx_1 \int_0^{x_1} \delta(x_2- {\cal H}^2(x_0) )dx_2\int_{x_2}^{x_1} \delta(x_3-{\cal H}^3(x_0) )dx_3 \nonumber
\end{align}
\end{widetext}
the first integral on the right gives the following conditions:\\
${\cal H}^3(x_0)> {\cal H}^2(x_0)$\\
${\cal H}^2(x_0)< {\cal H}(x_0)$\\
${\cal H}(x_0)> x_0$\\
which are satisfied for $x_0 \in [1/2,3/4]$. The second integral gives:\\
${\cal H}^2(x_0) < {\cal H}^3(x_0)< {\cal H}(x_0)$\\
${\cal H}^2(x_0)< {\cal H}(x_0)$\\
${\cal H}(x_0)< x_0$\\
which are satisfied for $x_0 \in [1/2+\sqrt{3}/4,1]$. Thus
 
$$\mathbb{P}^4_4 =\frac{1}{\pi} \left[ B_{\left[\frac{1}{2},\frac{3}{4}\right]}\left(\frac{1}{2},\frac{1}{2}\right)+B_{\left[\frac{1}{2}+\frac{\sqrt{3}}{4},1\right]}\left(\frac{1}{2},\frac{1}{2}\right)\right] =8/24.$$

\item $\mathbb{P}^4_5$

\begin{eqnarray}
&&\mathbb{P}^4_5 = \int_0^1 f(x_0)dx_0 \int_0^{x_0} \delta (x_1 - {\cal H}(x_0)) dx_1 \int_{x_1}^{x_0} \delta(x_2- {\cal H}^2(x_0) )dx_2\int_{x_2}^1 \delta(x_3-{\cal H}^3(x_0) )dx_3 \nonumber 
\end{eqnarray}
gives the following conditions:\\
${\cal H}^3(x_0)> {\cal H}^2(x_0)$\\
${\cal H}(x_0)<{\cal H}^2(x_0)< x_0$\\
which are satisfied for $x_0 \in [1/4+\sqrt{3}/4,1]$ and
 
$$\mathbb{P}^4_5 =\frac{1}{\pi} B_{\left[\frac{1}{4}+\frac{\sqrt{3}}{4},1\right]}\left(\frac{1}{2},\frac{1}{2}\right)=\frac{1}{6}\quad(=4/24).$$

\item $\mathbb{P}^4_6$

\begin{eqnarray}
&&\mathbb{P}^4_6 = \int_0^1 f(x_0)dx_0 \int_0^{x_0} \delta (x_1 - {\cal H}(x_0)) dx_1 \int_0^{x_1} \delta(x_2- {\cal H}^2(x_0) )dx_2\int_{x_1}^1 \delta(x_3-{\cal H}^3(x_0) )dx_3 \nonumber 
\end{eqnarray}
gives the following conditions:\\
${\cal H}^3(x_0)> {\cal H}(x_0)>{\cal H}^2(x_0)$\\
${\cal H}(x_0)< x_0$\\
which are never satisfied for the ${\cal H}(x)$ map (this is indeed based on the fact that the pattern $x_i>x_{i+1}<x_{i+2}$ is indeed a forbidden pattern in the orbit of ${\cal H}(x)$.Hence 
 
$$\mathbb{P}^4_6 =0.$$

\end{itemize}

\section*{APPENDIX II: Motif profiles for all subjects in the empirical study}

\begin{figure*}
\centering
\includegraphics[width=18.5 cm]{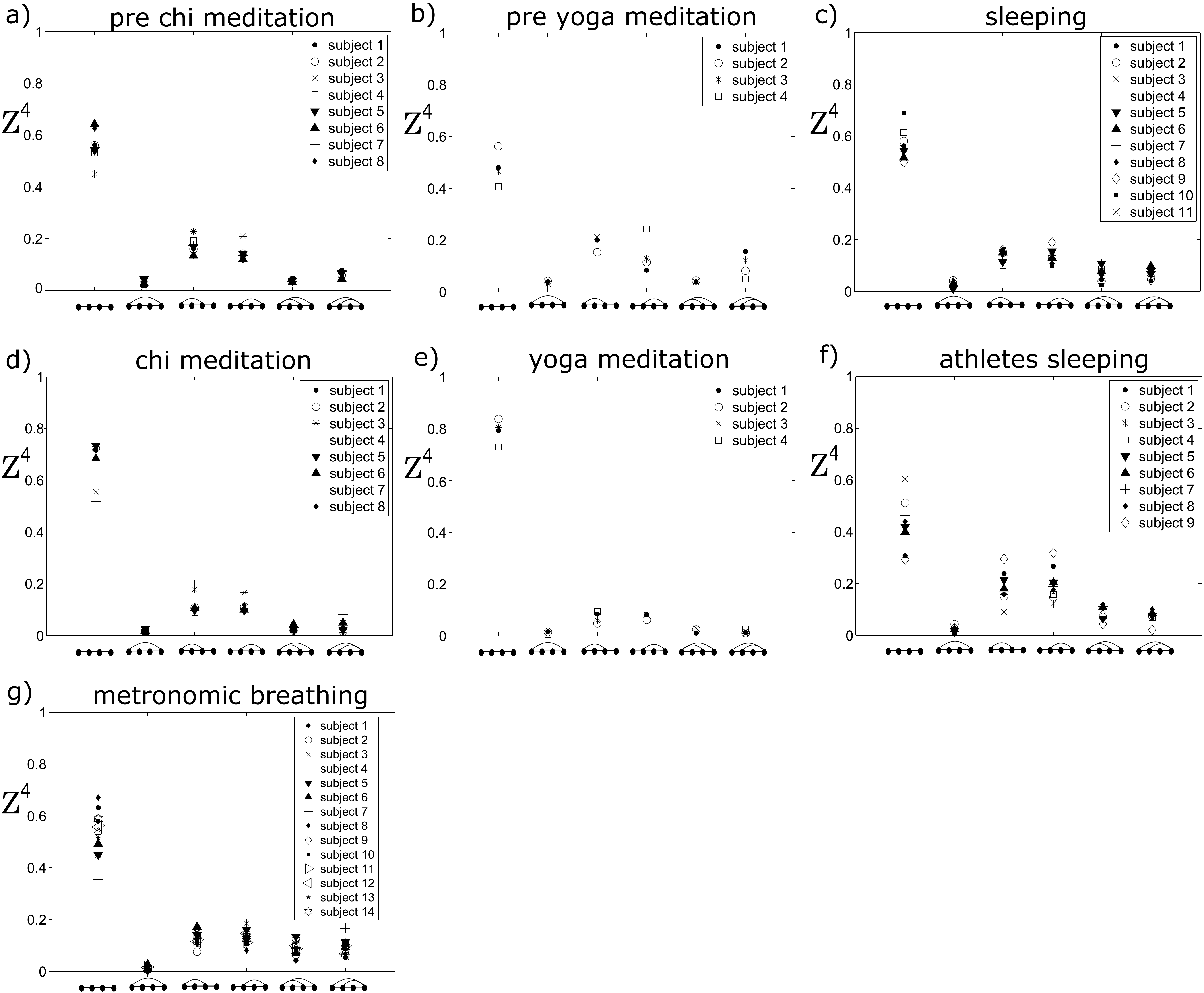}
\caption{HVG motif significance profile ${\bf Z}^4$ obtained by analysing heart rate time series form different groups of subjects in different states: a) 8 Chi meditators before the meditation practice; b) 4 Yoga meditators before the meditation practice; c) 11 subjects during sleeping; d) same 8 Chi meditators of a) during the meditation practice; e) same 4 Yoga meditators of b) during the meditation practice; f) 9 elite athletes during sleeping; ; g) 14 subjects during metronomic breathing at 0.25 Hz.}   
\label{fig:5}
\end{figure*}

In Figure \ref{fig:5} we give an overview of the 4-node motif profiles, measured for the different subjects in the different states. Interestingly, the motif that shows more variability in each of the given states is the one related to the type-1 motif, which we have seen to play a minor role in the case of the chaotic dynamics polluted with noise.\\

\subsection{Scores}

 The scores of the two components in terms of motifs are reported in Table \ref{tab:scores1}, and as expected the highest contribution to the first component  (0.874) is given by motif of type 1.

\begin{table*}[]
\centering
\label{tab:scores1}
\begin{tabular}{lllll}
\multicolumn{1}{c}{{\bf }} & \multicolumn{1}{c}{Yoga}    & \multicolumn{1}{c}{{\bf }}  & Chi                         &                             \\
                           & First Component                & Second Component & First Component                & Second Component                \\ \cline{2-5} 
\multicolumn{1}{l|}{\begin{minipage}{.1\textwidth}
\includegraphics[width= 0.7\textwidth]{4motif-1.pdf}
\end{minipage}}    & \multicolumn{1}{l|}{0.874}  & \multicolumn{1}{l|}{0.0346} & \multicolumn{1}{l|}{0.871}  & \multicolumn{1}{l|}{0.171}  \\ \cline{2-5} 
\multicolumn{1}{l|}{\begin{minipage}{.1\textwidth}
\includegraphics[width= 0.7\textwidth]{4motif-2.pdf}
\end{minipage}}    & \multicolumn{1}{l|}{-0.029} & \multicolumn{1}{l|}{-0.203} & \multicolumn{1}{l|}{-0.023} & \multicolumn{1}{l|}{-0.239} \\ \cline{2-5} 
\multicolumn{1}{l|}{\begin{minipage}{.1\textwidth}
\includegraphics[width= 0.7\textwidth]{4motif-3.pdf}
\end{minipage}}    & \multicolumn{1}{l|}{-0.379} & \multicolumn{1}{l|}{0.074}  & \multicolumn{1}{l|}{-0.376} & \multicolumn{1}{l|}{0.207}  \\ \cline{2-5} 
\multicolumn{1}{l|}{\begin{minipage}{.1\textwidth}
\includegraphics[width= 0.7\textwidth]{4motif-4.pdf}
\end{minipage}}    & \multicolumn{1}{l|}{-0.204} & \multicolumn{1}{l|}{0.731}  & \multicolumn{1}{l|}{-0.272} & \multicolumn{1}{l|}{0.666}  \\ \cline{2-5} 
\multicolumn{1}{l|}{\begin{minipage}{.1\textwidth}
\includegraphics[width= 0.7\textwidth]{4motif-5.pdf}
\end{minipage}}    & \multicolumn{1}{l|}{-0.043} & \multicolumn{1}{l|}{0.01}   & \multicolumn{1}{l|}{-0.048} & \multicolumn{1}{l|}{-0.173} \\ \cline{2-5} 
\multicolumn{1}{l|}{\begin{minipage}{.1\textwidth}
\includegraphics[width= 0.7\textwidth]{4motif-6.pdf}
\end{minipage}}    & \multicolumn{1}{l|}{-0.219} & \multicolumn{1}{l|}{-0.647} & \multicolumn{1}{l|}{-0.152} & \multicolumn{1}{l|}{-0.631} \\ \cline{2-5} 
\end{tabular}
\caption{Principal component scores obtained from PCA considering the Yoga meditators data subset (left) and the Chi meditators data subset (right).}
\end{table*}

\begin{table*}[]
\centering
\label{tab:scores2}\begin{tabular}{lllll}
\multicolumn{1}{c}{{\bf }} & Chi\&Yoga                   &                             & All States                  &                             \\
                           & First Component                & Second Component & First Component                & Second Component               \\ \cline{2-5} 
\multicolumn{1}{l|}{\begin{minipage}{.1\textwidth}
\includegraphics[width= 0.7\textwidth]{4motif-1.pdf}
\end{minipage}}    & \multicolumn{1}{l|}{0.874}  & \multicolumn{1}{l|}{0.08}   & \multicolumn{1}{l|}{0.881}  & \multicolumn{1}{l|}{0.133}  \\ \cline{2-5} 
\multicolumn{1}{l|}{\begin{minipage}{.1\textwidth}
\includegraphics[width= 0.7\textwidth]{4motif-2.pdf}
\end{minipage}}    & \multicolumn{1}{l|}{-0.027} & \multicolumn{1}{l|}{-0.181} & \multicolumn{1}{l|}{0.007}  & \multicolumn{1}{l|}{0.073}  \\ \cline{2-5} 
\multicolumn{1}{l|}{\begin{minipage}{.1\textwidth}
\includegraphics[width= 0.7\textwidth]{4motif-3.pdf}
\end{minipage}}    & \multicolumn{1}{l|}{-0.378} & \multicolumn{1}{l|}{0.089}  & \multicolumn{1}{l|}{-0.315} & \multicolumn{1}{l|}{0.477}  \\ \cline{2-5} 
\multicolumn{1}{l|}{\begin{minipage}{.1\textwidth}
\includegraphics[width= 0.7\textwidth]{4motif-4.pdf}
\end{minipage}}    & \multicolumn{1}{l|}{-0.235} & \multicolumn{1}{l|}{0.714}  & \multicolumn{1}{l|}{-0.294} & \multicolumn{1}{l|}{0.4}    \\ \cline{2-5} 
\multicolumn{1}{l|}{\begin{minipage}{.1\textwidth}
\includegraphics[width= 0.7\textwidth]{4motif-5.pdf}
\end{minipage}}    & \multicolumn{1}{l|}{-0.045} & \multicolumn{1}{l|}{-0.039} & \multicolumn{1}{l|}{-0.13}  & \multicolumn{1}{l|}{-0.58}  \\ \cline{2-5} 
\multicolumn{1}{l|}{\begin{minipage}{.1\textwidth}
\includegraphics[width= 0.7\textwidth]{4motif-6.pdf}
\end{minipage}}    & \multicolumn{1}{l|}{-0.188} & \multicolumn{1}{l|}{-0.664} & \multicolumn{1}{l|}{-0.15}  & \multicolumn{1}{l|}{-0.503} \\ \cline{2-5} 
\end{tabular}
\caption{Principal component scores obtained from PCA considering the subset data of Yoga and Chi meditators together (left) and and considering all data set (right).}
\end{table*}

\acknowledgments{We thank two anonymous referees for their helpful comments and suggestions.}

 {}

\end{document}